\documentclass{nature}
\usepackage[utf8]{inputenc}
\usepackage{graphicx}
\usepackage{graphics}
\graphicspath{ {./Figures/} }
\usepackage{duckuments} %
\usepackage{float}
\usepackage{lineno}
\usepackage{comment}
\usepackage{tabularx}
\usepackage{booktabs}
\usepackage{multirow}
\usepackage{amsmath,amssymb}
\usepackage{bbold}
\usepackage[dvipsnames]{xcolor}
\usepackage{mathtools}
\usepackage{braket}

\usepackage{textcomp,gensymb}
\usepackage{prettyref}
\newrefformat{fig}{Fig.~\ref{#1}}
\newrefformat{exfig}{Extended Data Fig.~\ref{#1}}

\usepackage{hyperref}
\hypersetup{
    colorlinks=true,
    linkcolor=blue,
    filecolor=magenta,      
    urlcolor=cyan,
    citecolor=red,
    pdftitle={Spin-Orbit Enhanced Superconductivity in Bernal Bilayer Graphene},
}

\usepackage{caption}
\DeclareCaptionLabelFormat{vertline}{#1 #2 $|$}
\usepackage{enumitem}

\makeatletter \let\saved@includegraphics\includegraphics
\AtBeginDocument{\let\includegraphics\saved@includegraphics} 
\renewenvironment{figure}{\@float{figure}}{\end@float}
\renewenvironment{table}{\@float{table}[p]\sffamily\fontsize{7pt}{12pt}\selectfont}{\end@float} %
\makeatother

\definecolor{orange}{rgb}{1,0.5,0}

\newcommand{\beginsupplement}{%
        \setcounter{table}{0}
        \setcounter{figure}{0}
        \renewcommand{\figurename}{Extended Data Fig.}
        \renewcommand{\tablename}{SI Table}
     }

\newcommand{\vk}{{\boldsymbol{k}}}
\newcommand{\vK}{{\boldsymbol{K}}}
\newcommand{\vq}{{\boldsymbol{q}}}
\newcommand{\vr}{{\boldsymbol{r}}}
\newcommand{\vb}{{\boldsymbol{b}}}

\newcommand{\vA}{{\boldsymbol{A}}}
\newcommand{\vB}{{\boldsymbol{B}}}
\newcommand{\vs}{{\boldsymbol{s}}}

\renewcommand{\vec}[1]{{\boldsymbol #1}} %

\DeclareMathAlphabet{\mathcald}{U}{dutchcal}{m}{n}
\SetMathAlphabet{\mathcald}{bold}{U}{dutchcal}{b}{n}
\DeclareMathAlphabet{\mathalt}{U}{dutchcal}{b}{n}

\title{Spin-Orbit Enhanced Superconductivity in Bernal Bilayer Graphene}
\author{Yiran Zhang$^{1,2,3}$, Robert Polski$^{1,2}$, 
Alex Thomson$^{2,3,4}$, Étienne Lantagne-Hurtubise$^{2,3}$, Cyprian Lewandowski$^{2,3}$,  Haoxin Zhou$^{1,2}$, Kenji Watanabe$^5$, Takashi Taniguchi$^5$, 
Jason Alicea$^{2,3}$, and Stevan Nadj-Perge$^{1,2\dagger}$}
\begin{document}

\maketitle

\begin{affiliations}

  \item T. J. Watson Laboratory of Applied Physics, California Institute of
  Technology, 1200 East California Boulevard, Pasadena, California 91125, USA
  \item Institute for Quantum Information and Matter, California Institute of
  Technology, Pasadena, California 91125, USA
  \item Department of Physics, California Institute of Technology, Pasadena,
  California 91125, USA
  \item Department of Physics, University of California, Davis, California 95616, USA
  \item National Institute for Materials Science, Namiki 1-1, Tsukuba, Ibaraki
  305 0044, Japan
  \item[$^\dagger$] Correspondence: s.nadj-perge@caltech.edu
  
\end{affiliations}
\begin{abstract}

In the presence of a large perpendicular electric field, Bernal-stacked bilayer
graphene (BLG) features several broken-symmetry metallic
phases\cite{zhouIsospinMagnetismSpinpolarized2022,
delabarreraCascadeIsospinPhase2021, seilerQuantumCascadeNew2021} as well as
magnetic-field-induced
super-conductivity\cite{zhouIsospinMagnetismSpinpolarized2022}. The
superconducting state is quite fragile, however, appearing only in a narrow
window of density and with a maximum critical temperature
$\mathbf{T_c\approx30}$~mK. Here, we show that placing monolayer tungsten
diselenide (WSe$\mathbf{_{2}}$) on BLG promotes Cooper pairing to an
extraordinary degree: superconductivity appears at zero magnetic field,
exhibits an order of magnitude enhancement in $\mathbf{T_c}$, and occurs over a
density range that is wider by a factor of eight. By mapping quantum
oscillations in BLG-WSe$\mathbf{_2}$ as a function of electric field and
doping, we establish that superconductivity emerges throughout a region whose
normal state is polarized, with two out of four spin-valley flavours
predominantly populated. In-plane magnetic field measurements further reveal a
striking dependence of the critical field on doping, with the
Chandrasekhar-Clogston (Pauli) limit roughly obeyed on one end of the
superconducting dome yet sharply violated on the other. Moreover, the
superconductivity arises only for perpendicular electric fields that push BLG
hole wavefunctions towards WSe$_2$---suggesting that proximity-induced (Ising)
spin-orbit coupling plays a key role in enhancing the pairing. Our results pave
the way for engineering robust, highly tunable, and ultra-clean graphene-based
superconductors.
 
\end{abstract}

Strong interactions between electrons often lead to a rich competition of
symmetry-breaking phases throughout the parameter space. This competition can
be significantly altered by external perturbations that lower the energy for
one of the phases at the expense of the others. One recent example of such a
phase diagram modification occurs in magic-angle twisted bilayer
graphene\cite{bistritzerMoireBandsTwisted2011} aligned with hexagonal boron
nitride (hBN), where sublattice polarization stabilizes a Chern insulating
phase near a filling of three electrons per moir\'e unit cell at the expense of
suppressing
superconductivity\cite{sharpeEmergentFerromagnetismThreequarters2019,
serlinIntrinsicQuantizedAnomalous2019}. Here we investigate the symmetry-broken
phases in Bernal-stacked bilayer graphene (BLG) coupled to a WSe$_2$ monolayer
and show that the phase diagram is altered such that superconductivity is
strongly enhanced.

\begin{figure}[p]
    \centering
    \includegraphics[width=16cm]{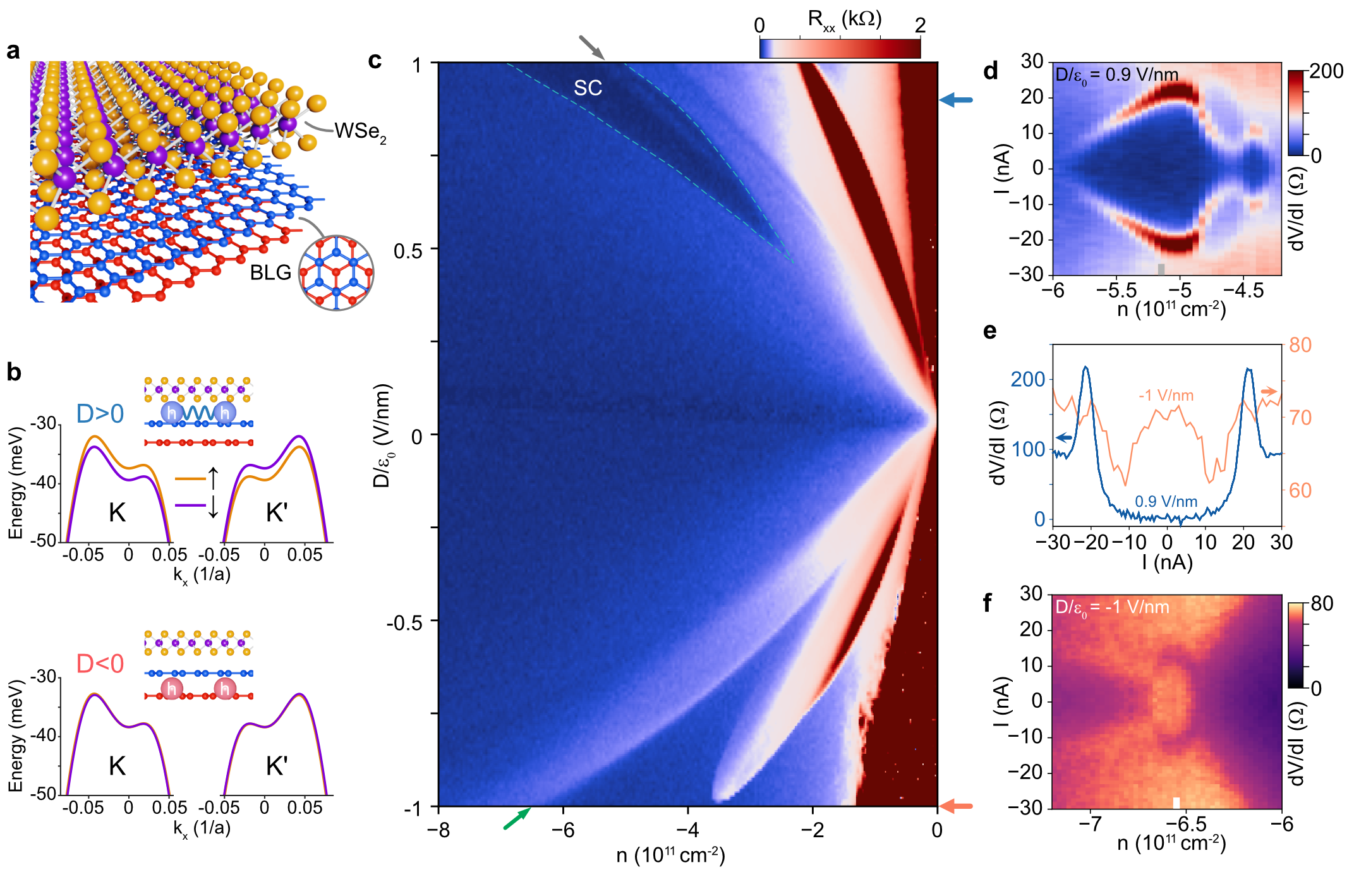}
    \caption{{\bf Phase diagram of BLG-WSe$_2$ and superconductivity at zero magnetic field.} {\bf a}, Schematic of a BLG-WSe$_2$ structure showing the crystal lattice of Bernal-stacked bilayer graphene (blue and red) and a WSe$_2$ monolayer (yellow and purple) on top. {\bf b}, Non-interacting valence bands near the $K$ and $K'$ points of the Brillouin zone for $D/\epsilon_0 = 1~\text{V/nm}$ (top) and $-1~\text{V/nm}$ (bottom), calculated by including an Ising SOC ($\lambda_I = 1$~meV) on the top layer. Schematics show that when BLG is hole-doped, electronic wavefunctions are polarized towards the top layer for $D > 0$, and towards the bottom layer for $D < 0$. {\bf c}, $R_{xx}$ versus doping density $n$ and displacement field $D$ measured at zero magnetic field. Flavour-polarized states show strong asymmetry with respect to the sign of $D$ field. Superconductivity (delineated by a dashed line)
    spans across wide doping and $D$ ranges at positive $D$ fields (wavefunctions are strongly polarized towards the WSe$_2$). A competing resistive phase appears
    in the middle of the superconducting region, as marked by the grey arrow. {\bf d},{\bf f}, $dV/dI$
    versus $n$ and bias current $I$ measured at $D/\epsilon_0 = 0.9~\text{V/nm}$ ({\bf d}) and $-1~\text{V/nm}$ ({\bf f}), respectively. {\bf e}, Blue and orange curves are line cuts from {\bf d} and {\bf f}, respectively, with the densities marked by the coloured bars.}
    \label{fig:Fig1}
\end{figure}

Figure \ref{fig:Fig1}a shows the BLG-WSe$_2$ stack while
\prettyref{fig:Fig1}b displays the non-interacting electronic bands of BLG in
the presence of a perpendicular electric displacement field ($D$). In a finite
$D$ field, BLG features a band gap at charge
neutrality\cite{mccannAsymmetryGapElectronic2006,
zhangDirectObservationWidely2009} as well as trigonal
warping\cite{mccannElectronicPropertiesBilayer2013} and prominent Van Hove
singularities (VHS) near the very weakly dispersive band edge. Due to the large
density of states, interactions between electrons are greatly amplified when
the chemical potential crosses the VHS. Additionally, a finite $D$ field
significantly polarizes the low-energy electronic
wavefunctions\cite{mccannAsymmetryGapElectronic2006,
mccannElectronicPropertiesBilayer2013} (\prettyref{fig:Fig1}b insets)
towards the top or bottom layers and on different sublattices $A$ and $B$. When
combined with WSe$_2$ placed on one side, BLG becomes an ideal experimental
platform for probing the interplay between electronic
correlations\cite{zhouIsospinMagnetismSpinpolarized2022,
delabarreraCascadeIsospinPhase2021, seilerQuantumCascadeNew2021} and induced
spin-orbit coupling (SOC)\cite{wangOriginMagnitudeDesigner2016,
gmitraProximityEffectsBilayer2017, khooOnDemandSpinOrbit2017,
khooTunableQuantumHall2018, islandSpinOrbitdrivenBand2019,
wangQuantumHallEffect2019, liTwistangleDependenceProximity2019}.
  
Longitudinal resistance $R_{xx}$ measured as a function of carrier density $n$
and $D$ at zero magnetic field shows peaks or dips that emerge and separate
from each other as $|D|$ is increased (\prettyref{fig:Fig1}c). These features
can be associated with an interplay of Lifshitz transitions and breaking of
spin and valley symmetries, similar to the case of hBN-encapsulated
BLG\cite{zhouIsospinMagnetismSpinpolarized2022}. Importantly, the resulting
phase diagram is strongly asymmetric with respect to the sign of $D$ field.
Focusing on hole doping, for both signs of $D$, the largest resistance peaks
(red diagonal regions in \prettyref{fig:Fig1}c) correspond to phases that
possess a single spin-valley flavour-polarized Fermi surface, which we denote
as $\mathrm{FP}(1)_{\pm}$ ($\mathrm{FP}(n)$ denotes a flavour-polarized phase
with $n$ degenerate Fermi pockets and $\pm$ denotes the sign of $D$; see
\prettyref{exfig:QO_cuts} for the identification of spin-valley degeneracy
though quantum oscillations). For positive $D$, this resistive feature spans
beyond $D/\epsilon_0=+1$~V/nm but is suppressed by $D/\epsilon_0=-0.75$~V/nm
for negative $D$.
  
The pronounced $\pm D$ asymmetry highlights the role of Ising SOC in defining
the phase diagram of BLG-WSe$_2$. Theoretical
calculations\cite{gmitraProximityEffectsBilayer2017, khooOnDemandSpinOrbit2017}
(\prettyref{fig:Fig1}b) confirm that Ising SOC is induced only on the top layer
proximate to WSe$_2$ and that, correspondingly, the SOC-induced spin splitting
in the valence band is largely restricted to $D > 0$---consistent with the
$D$-asymmetric experimental data (\prettyref{fig:Fig1}c; see also Methods). In
contrast, Rashba SOC is expected to couple symmetrically to the valence and
conduction bands due to their sublattice polarization, and thus cannot account
for the pronounced asymmetry between $\pm D$ (see Supplementary Information
(SI), section \ref{theory: continuum_model_BLG} for further discussion).
  
The most striking difference in the BLG-WSe$_2$ phase diagram between positive
and negative $D$ fields is the emergence of a broad zero-resistance region
corresponding to superconductivity at $D>0$. No analogous region has been
observed in hBN-encapsulated BLG, where superconductivity only appears in a
finite in-plane magnetic field\cite{zhouIsospinMagnetismSpinpolarized2022}. The
critical current of the zero-magnetic-field superconductivity in BLG-WSe$_2$
exhibits nontrivial doping dependence (\prettyref{fig:Fig1}d,e), with two
distinct maxima (the larger of which reaches $20$~nA). By contrast, at $D<0$ a
different phase (\prettyref{fig:Fig1}e,f) exhibiting highly nonlinear
current-dependent resistance is observed for similar values of $n$ and $|D|$
(marked by a green arrow in \prettyref{fig:Fig1}c). This resistive phase is
suppressed by small magnetic fields and is similar to the zero-magnetic-field
phase that has been reported in hBN-encapsulated
BLG\cite{zhouIsospinMagnetismSpinpolarized2022}.

\begin{figure}[p]
    \centering
    \includegraphics[width=16cm]{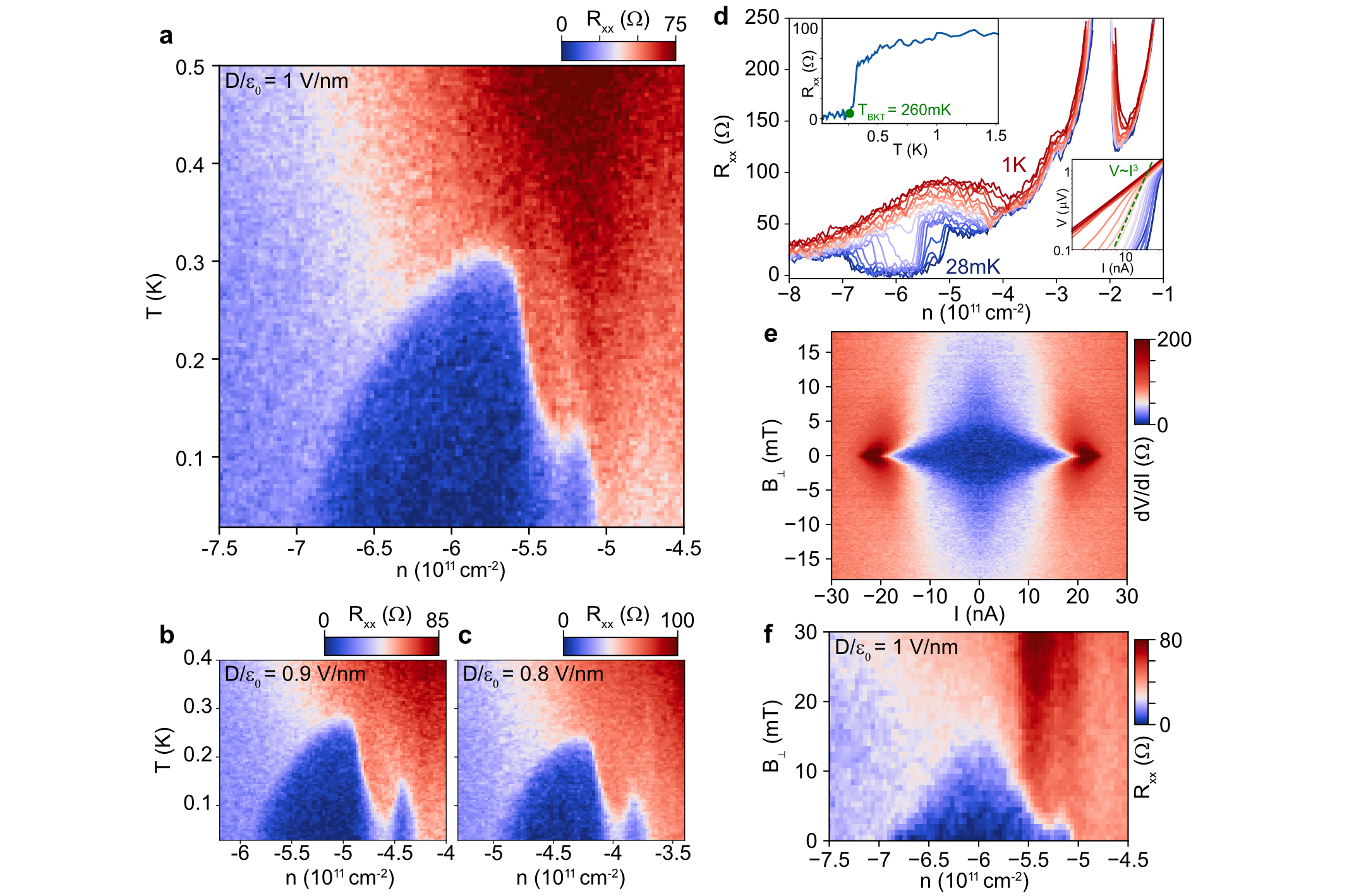}
    \caption{{\bf Evolution of the superconducting phase with temperature and out-of-plane magnetic field.} {\bf a}--{\bf c}, $R_{xx}$ versus density $n$ and temperature for hole doping, showing superconducting domes in the $\mathrm{FP}(2,2)_{+}$ phase for  $D/\epsilon_0 = 1~\text{V/nm}$ ({\bf a}), $0.9~\text{V/nm}$
      ({\bf b}), and $0.8~\text{V/nm}$ ({\bf c}), respectively. A competing resistive phase intersects the superconducting domes at these $D$ fields, which is also evident in the $B_{\perp}$ field dependence ({\bf f}). {\bf d}, Line cuts
      of $R_{xx}$ versus $n$ for a range of temperatures (from $28$~mK to $1$~K) measured at $D/\epsilon_0 = 1~\text{V/nm}$. The top inset is $R_{xx}$ versus temperature  measured at $n = -5.75\times 10^{11} ~\text{cm}^{-2}$ showing a superconducting transition. The bottom inset shows the corresponding $V$--$I$ plot at various temperatures. The green dashed line marks where $V\sim I^{3}$, from which we determine $T_{BKT} = 260$~mK. {\bf e}, Critical current disappearing with $B_{\perp}$ field measured at $D/\epsilon_0 = 0.9~\text{V/nm}$, $n = -5.05\times 10^{11} ~\text{cm}^{-2}$. {\bf f}, $R_{xx}$ versus $n$ and $B_\perp$ field around the superconducting region for $D/\epsilon_0 = 1~\text{V/nm}$.
      }
    \label{fig:Fig2}
\end{figure}

The evolution of critical temperature $T_c$ with doping and displacement field
provides further insights into the nature of the superconductivity
(\prettyref{fig:Fig2}a-c). The superconducting dome occupies a wide range of
doping ($\sim 2\times 10^{11} ~\text{cm}^{-2}$; see also \prettyref{fig:Fig1}c)
and features a maximal $T_c$ of approximately $300$~mK. Figure \ref{fig:Fig2}d shows
$R_{xx}$ line cuts at different temperatures; insets show nonlinear $I$--$V$
curves at optimal doping, yielding a Berezinskii–Kosterlitz–Thouless (BKT)
transition temperature $T_{BKT} \approx 260$~mK (estimated by the temperature
where $V\sim I^3$). We emphasize that the superconducting critical temperature
observed here is an order of magnitude larger than the $T_{c}$ in
hBN-encapsulated BLG measured at optimal in-plane magnetic field. Moreover, the
relatively high $T_{c}$ does not appear to be sensitive to minor changes of $D$
field, further substantiating the robustness of the superconducting phase.
Figure \ref{fig:Fig2}e,f shows the evolution of the superconducting phase in
the presence of an out-of-plane magnetic field $B_{\perp}$. The maximal
critical field $B_{c\perp} \approx 15$~mT at base temperature yields a
corresponding Ginzburg-Landau coherence length $\xi_\mathrm{GL} =
\sqrt{\Phi_0/(2\pi B_{c\perp})} \approx 150$~nm ($\Phi_0$ is the superconductor
flux quantum), while the mean free path $\ell_{mf}$ of BLG-WSe$_{2}$ is around
$10$ $\mu$m (see Methods and \prettyref{exfig:magnetic focusing}).
Superconductivity thus resides deep in the clean limit,
$\xi_\mathrm{GL}/\ell_{mf} < 0.02$, similar to the case of hBN-encapsulated
Bernal bilayer and rhombohedral trilayer
graphene\cite{zhouIsospinMagnetismSpinpolarized2022,
zhouSuperconductivityRhombohedralTrilayer2021}.
    
Another prominent feature of both the $T$ and $B_{\perp}$ field dependence
(\prettyref{fig:Fig2}a-c and f) is a resistive peak that intersects the
superconducting dome, effectively splitting it into two regions within a
certain range of $D$ fields (marked by a grey arrow in \prettyref{fig:Fig1}c).
This peak signals the presence of another phase that appears to compete with
superconductivity. Both the doping range where this state occurs and its
disappearance at relatively low magnetic fields are features shared by the
resistive phase observed for $D<0$ (see the green arrow in
\prettyref{fig:Fig1}c) and in hBN-encapsulated
BLG\cite{zhouIsospinMagnetismSpinpolarized2022}. Moreover, both the resistive
peak and superconductivity feature a broken-symmetry parent state with two
large and emerging small Fermi pockets (see discussion below), suggesting
that transport in this region is highly sensitive to the exact details of the
spin-valley ground states (see \prettyref{exfig:theory_figure_cascade} and SI,
section \ref{theory: Polarized_phases} for possible competition between the
ground states).

\begin{figure}[p]
    \centering
    \includegraphics[width=16cm]{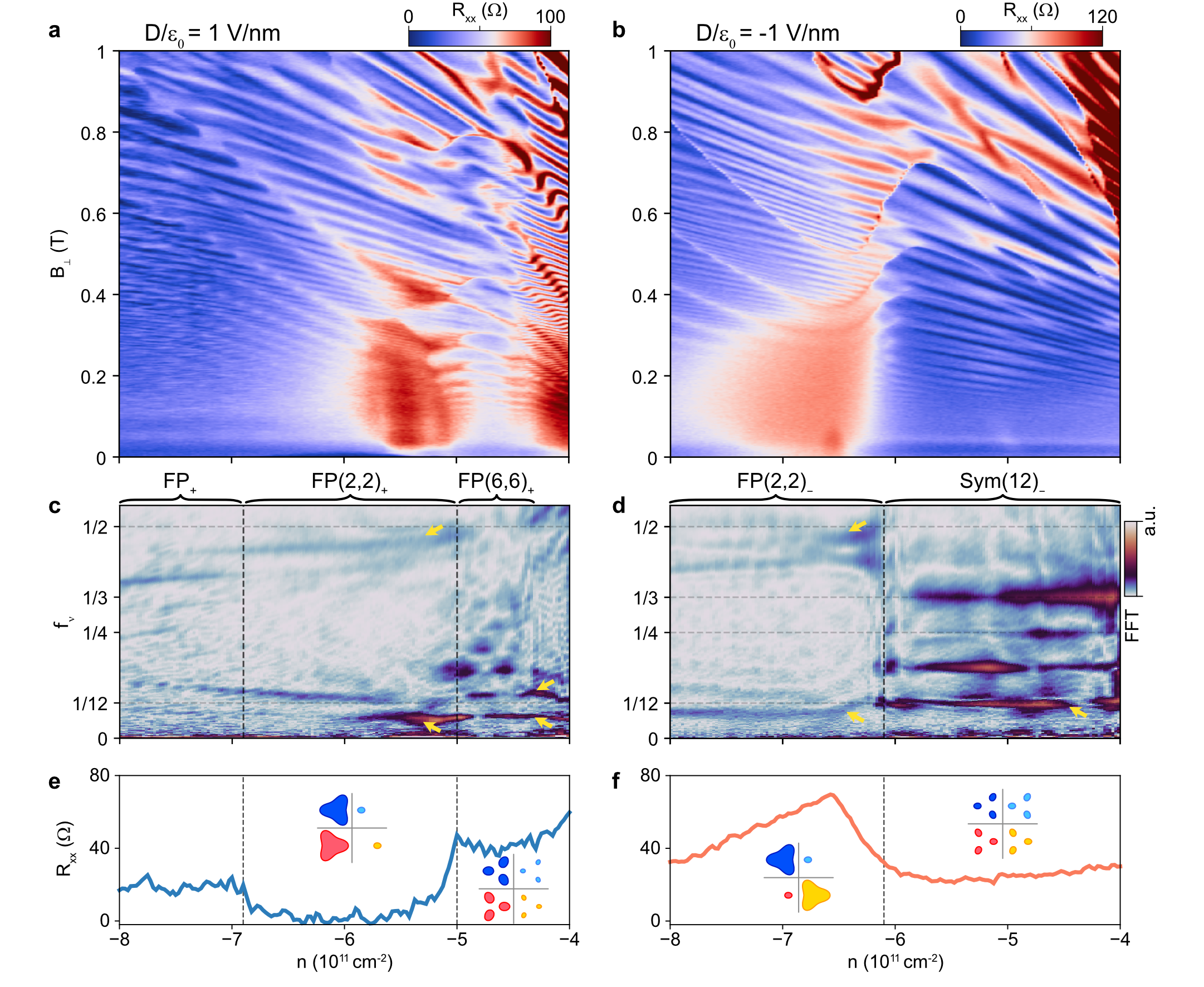}
    \caption{{\bf Fermi surface structure of the superconducting state.} {\bf a},{\bf b}, $R_{xx}$ versus out-of-plane magnetic field $B_\perp$ and doping density $n$ for $D/\epsilon_0 = 1~\text{V/nm}$ ({\bf a}) and $-1~\text{V/nm}$ ({\bf b}), respectively. {\bf c},{\bf d}, Fast Fourier transform (FFT) of $R_{xx}(1/B_\perp)$ versus $n$ and $f_\nu$, where $f_\nu$ reveals the fraction of the total Fermi surface area enclosed by a cyclotron orbit. The relevant FFT peaks are marked by yellow arrows.
    FFT in {\bf c} and {\bf d} is converted from the $R_{xx}$ data within $0.05$~T~$<B_\perp<0.6$~T in {\bf a} and {\bf b}, respectively. {\bf e},{\bf f}, Line cuts of $R_{xx}$ versus $n$ showing superconductivity for $D/\epsilon_0 = 1~\text{V/nm}$ ({\bf e}) and resistive phase for $D/\epsilon_0 = -1~\text{V/nm}$ ({\bf f}) at $B_\perp = 0$~T. Schematics depict the possible Fermi surface structures for the different phases given that spin-valley flavours are not mixed.}
    \label{fig:Fig3}
\end{figure}

The $D$-field asymmetry is further highlighted by low field 
($B_{\perp}<1~\text{T}$) quantum oscillations measured at 
$D/\epsilon_0 = 1~\text{V/nm}$ and $-1~\text{V/nm}$, which imply distinct Fermi 
surface structures within the superconductivity region for $D>0$ 
(\prettyref{fig:Fig3}a,c,e) and within the resistive phase for $D<0$ 
(\prettyref{fig:Fig3}b,d,f). Fourier transforms 
of the oscillations---taken with respect to $1/B_{\perp}$---reveal the 
phases in the relevant doping ranges. To resolve the relative sizes 
of the Fermi pockets of the different flavour-polarized phases, 
the Fourier transform of $R_{xx}(1/B_{\perp})$ is normalized by the 
frequency corresponding to the full doping density, 
$f_\mathrm{norm} = n\times h/e$, so that the resulting frequency 
$f_{\nu}$ reveals the fraction of the total Fermi 
surface area enclosed by a cyclotron orbit (\prettyref{fig:Fig3}c,d). 

At $D/\epsilon_0 = -1~\text{V/nm}$, the resulting phase diagram is remarkably
similar to that reported on hBN-encapsulated BLG without
WSe$_2$\cite{zhouIsospinMagnetismSpinpolarized2022} (see also
\prettyref{exfig:negativeDfield}). In addition to the zero-field resistive
phase discussed before (\prettyref{fig:Fig1}f), at low densities ($|n| <
6\times 10^{11}~\text{cm}^{-2}$) we observe a Fourier transform peak at
$f_{\nu}=1/12$ (along with its higher harmonics) corresponding to a spin-valley
symmetric phase with $12$ degenerate Fermi pockets produced by trigonal warping
(denoted as $\mathrm{Sym}(12)_{-}$). Upon further hole doping, BLG transitions
into another phase with two frequency peaks at $f_{\nu}^{(1)} < 1/2$ and
$f_{\nu}^{(2)} < 1/12$ such that $f_{\nu}^{(1)}+f_{\nu}^{(2)} = 1/2$. This
phase can be identified as a spin-valley flavour-polarized phase---denoted
$\mathrm{FP}(2,2)_{-}$---with two majority ($f_{\nu}^{(1)} < 1/2$) and two
minority ($f_{\nu}^{(2)} < 1/12$) flavours. The resemblance between our $D<0$
data and hBN-encapsulated BLG\cite{zhouIsospinMagnetismSpinpolarized2022}
suggests that SOC does not play a major role for $D<0$.
 
\begin{figure}[p]
    \centering
    \includegraphics[width=16cm]{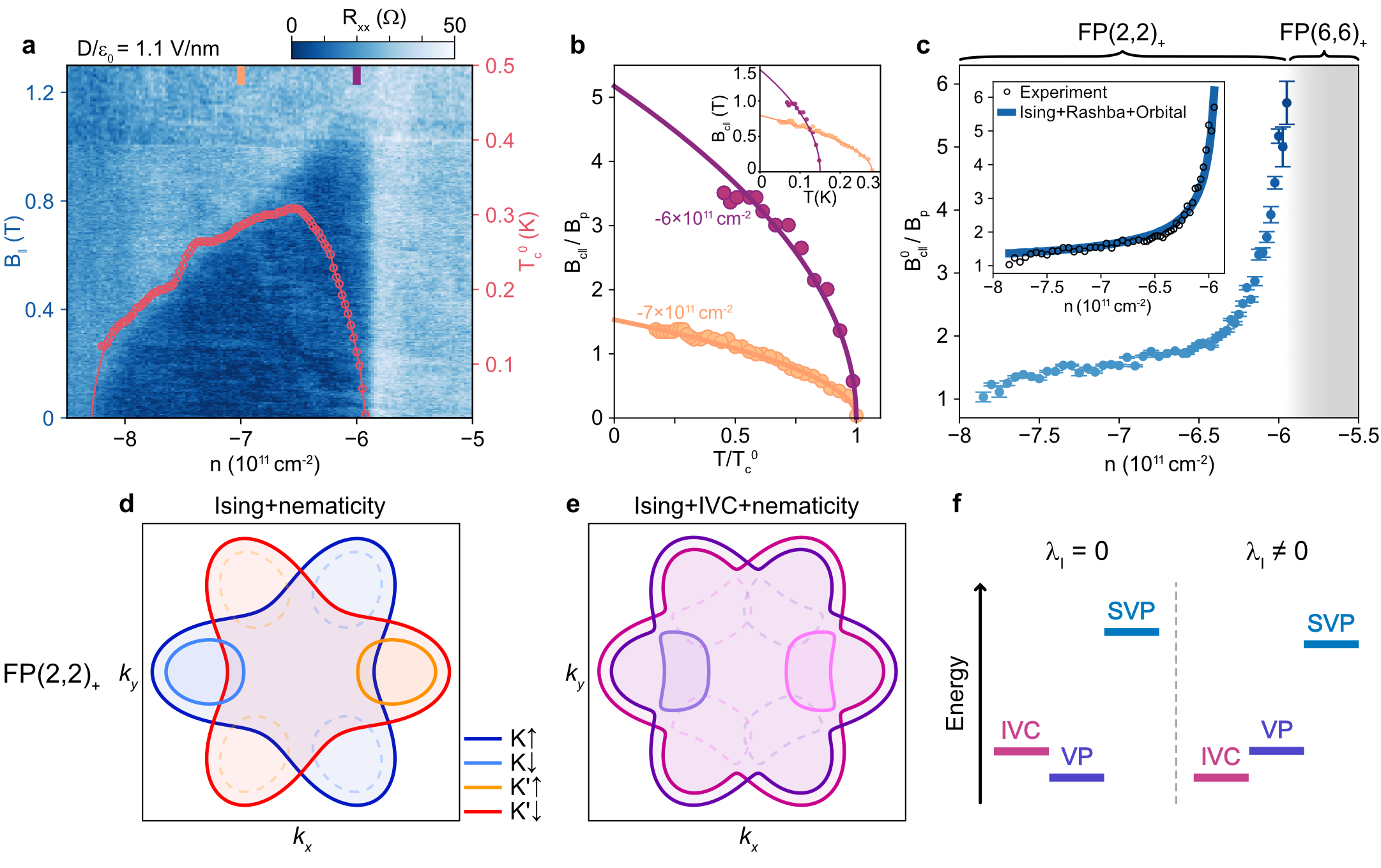}
    \caption{{\bf Doping-dependent Pauli-limit violation.} {\bf a}, $R_{xx}$
    versus in-plane magnetic field $B_\parallel$ and doping density $n$ for
    $D/\epsilon_0 = 1.1~\text{V/nm}$. The red line delineates the $T$-dependent
    superconducting dome, and open circles indicate the zero-magnetic-field
    critical temperature $T_{c}^{0}$ that is defined by the temperature at which
    $R_{xx}$ is $50\%$ of the normal state resistance.
    {\bf b}, The ratio of in-plane critical magnetic field $B_{c\parallel}$ (the
    field at which $R_{xx}$ is $50\%$ of the normal state resistance) to the
    Pauli-limit field $B_p = 1.86~\text{T}/\text{K}\times T_c^0$ is plotted as a
    function of normalized temperature $T/T_{c}^{0}$ at two doping densities $n
    = -6\times 10^{11} ~\text{cm}^{-2}$ and $-7\times 10^{11} ~\text{cm}^{-2}$.
    The data are well-fit by the phenomenological formula $T/T_{c}^{0} =
    1-(B_{c\parallel}/B_{c\parallel}^{0})^2$ (solid lines). Inset: the same data
    as in the main panel but plotted in $B_{c\parallel}$ versus $T$. {\bf c},
    Pauli violation ratio $B_{c\parallel}^{0}/B_p$ as a function of density $n$.
    Error bars are the standard deviation obtained from the phenomenological
    fitting. $B_{c\parallel}^{0}/B_p$ changes from roughly six to one as the
    doping is increased. The doping trend of the PVR is well captured by a model
    (blue line in the inset) taking into account fixed Ising SOC together with
    doping-dependent Rashba SOC and constant orbital depairing (further
    modelling is included in \prettyref{exfig:theory_figure_PLV} and SI,
    section~\ref{theory: PVR_fit}). {\bf d},{\bf e}, Fermi surfaces of the
    $\mathrm{FP}(2,2)_{+}$ phase with Ising SOC and nematic order ({\bf d}), or
    allowing for inter-valley coherent (IVC) order ({\bf e}). Dashed Fermi
    pockets correspond to the condition that nematic order is absent. {\bf f},
    Schematics of a proposed scenario where Ising SOC tilts the energy balance
    towards IVC order, within which the development of superconductivity is more
    favoured, at the expense of a state which is not conducive to pairing, e.g.,
    a valley polarized (VP) state.} 
    
    \label{fig:Fig4}
\end{figure}

At $D/\epsilon_0 = 1~\text{V/nm}$ (\prettyref{fig:Fig3}c,e), where the
wavefunctions are strongly polarized towards WSe$_2$, we see a few notable
differences (see \prettyref{exfig:Fan and FFT} for data at different $D$
fields). First, at low densities, one of the Fourier frequency peaks clearly
appears below $f_\nu=1/12$, suggesting the existence of Fermi surfaces whose
occupancy is smaller relative to $\mathrm{Sym}(12)_{-}$. As we can identify two
independent frequencies in this region, we denote this phase as
$\mathrm{FP}(6,6)_{+}$, with six bigger and six smaller Fermi pockets. Given
the lack of correlation signatures at the similar region for $D<0$, the
explicit flavour polarization here likely originates from spin-orbit induced
band splitting. Second, the transition between the $\mathrm{FP}(6,6)_{+}$ phase
and the adjacent $\mathrm{FP}(2,2)_{+}$ phase (with two big and two small Fermi
pockets) occurs at a lower hole density of $|n|=5\times
10^{11}~\text{cm}^{-2}$. Finally, we observe that superconductivity is
established throughout the $\mathrm{FP}(2,2)_{+}$ phase (except a small region
where it competes with the resistive phase) ending on the high doping side with
the onset of another complex flavour-polarized phase characterized by the
occurrence of additional frequency peaks (\prettyref{fig:Fig3}c,e; see also SI,
section \ref{theory: Polarized_phases} for the Fermi-surface candidates).
Importantly, in $\mathrm{FP}(2,2)_{+}$, as for $\mathrm{FP}(2,2)_{-}$, we find
that $f_{\nu}^{(1)}+f_{\nu}^{(2)} = 1/2$. Given the non-interacting band
structure of \prettyref{fig:Fig1}b, this observation implies that the carriers
in each minority flavour are spontaneously polarized to one of the trigonally
warped pockets---pointing towards nematic
order\cite{dongIsospinFerromagnetismMomentum2021, huangSpinOrbitalMetallic2022}
(\prettyref{fig:Fig4}d,e).

In-plane magnetic field measurements further illuminate the unconventional 
nature of superconductivity in BLG-WSe$_2$ (\prettyref{fig:Fig4} and \prettyref{exfig:inplanefield}). Figure~\ref{fig:Fig4}a shows 
$R_{xx}$ as a 
function of density $n$ and in-plane magnetic field $B_{\parallel}$ for the 
superconducting region (dark blue) at 
$D/\epsilon_0 = 1.1~\text{V/nm}$. When approaching the superconductivity from 
low densities $|n|$, the in-plane critical field $B_{c\parallel}$ quickly 
reaches a maximum near the phase boundary separating $\mathrm{FP}(2,2)_{+}$ and 
$\mathrm{FP}(6,6)_{+}$, and then slowly decreases with further hole doping. 
Conversely, the critical temperature measured at zero $B_\parallel$ field, $T_{c}^0$ (red open circles), shows a more symmetric dome shape with a 
maximum at higher $|n|$.
The interplay between $B_{c\parallel}$ and $T_{c}^0$ suggests that the violation 
of the Pauli limit ($B_p = 1.86~\text{T}/\text{K}\times T_c^0$ for a 
weak-coupling spin-singlet BCS superconductor with $g$-factor $g=2$) varies with
doping. As an example, \prettyref{fig:Fig4}b shows $B_{c\parallel}/B_p$ as a 
function of temperature ($T$ normalized to $T_c^0$) at two representative 
densities. Both curves are well-fit by the phenomenological
relation $T/T_{c}^{0} = 1-(B_{c\parallel}/B_{c\parallel}^{0})^2$ (solid lines;
$B_{c\parallel}^0$  denotes the critical field at zero temperature).
However, they show distinct Pauli violation ratios (PVR) 
$B_{c\parallel}^{0}/B_p$:
for high $|n|$ (orange curve, $n = -7\times 10^{11} ~\text{cm}^{-2}$), 
$B_{c\parallel}^{0}/B_p\approx1.5$ which is close to the ratio expected from 
weak coupling BCS theory.
The purple curve ($n = -6\times 10^{11} ~\text{cm}^{-2}$), however, shows 
$B_{c\parallel}^{0}/B_p\approx5$, strongly violating the Pauli limit.
Overall the PVR changes from roughly six to one as the doping is increased 
(\prettyref{fig:Fig4}c; consistent results are obtained by extracting 
$B^0_{c \parallel}$ at base temperature, see \prettyref{exfig:inplanefield}f). 
Note that the PVR values at the phase boundaries represent a lower limit due to 
possible imperfect in-plane alignment of the sample; see Methods for further 
discussion.

Among graphene-based superconductors, the striking gate-tunability of the PVR 
appears unique to BLG-WSe$_2$. In our BLG-WSe$_2$ device, the agreement between 
the coherence length data and a weak-coupling assumption 
(\prettyref{exfig:inplanefield}e inset) suggests 
that the variation of coupling strength is small and thus cannot 
explain the dramatic change in PVR.  Note also that even in moir\'e graphene, 
where superconductivity can be tuned from weak to strong 
coupling\cite{parkTunableStronglyCoupled2021, 
kimSpectroscopicSignaturesStrong2021}, the PVR is largely insensitive to  
doping\cite{caoPaulilimitViolationReentrant2021}. An alternative 
possibility is that the Fermi pockets responsible for superconductivity evolve 
non-trivially with doping in a manner that significantly alters the in-plane 
critical field. 

The large PVR of $B_{c \parallel}/B_p \sim 6$ on the low hole doping side of
the superconducting dome evokes the phenomenology of Ising superconductivity
observed in transition metal
dichalcogenides\cite{luEvidenceTwodimensionalIsing2015,
saitoSuperconductivityProtectedSpinvalley2016,
xiIsingPairingSuperconducting2016} (TMDs). Ising superconductivity refers to a
scenario in which pairing connects time-reversed states, e.g., $\ket{\vk,
\uparrow}$ and $\ket{-\vk, \downarrow}$, with spins oriented along a fixed
quantization axis selected by Ising SOC. Here $\lambda_I \approx
0.7~\text{meV}$---estimated from quantum Hall measurements at small $D$ (see
Methods and \prettyref{exfig:QHE_ising})---far exceeds $\Delta = 1.76 k_B T_c
\approx 0.02$ meV estimated from weak-coupling BCS scaling. The resulting
Cooper pairs enjoy resilience against in-plane fields that rotate the spins
away from this preferred axis, naturally leading to significant Pauli-limit
violation as measured on the low hole doping side of the dome. The substantial
PVR reduction on the high hole doping side is more puzzling and implies that
the ground state cannot evolve into a predominantly spin or spin-valley
polarized phase. This reduction could emerge from an interplay between a
doping-dependent change in the flavour polarization of the parent ${\rm
FP}(2,2)_+$ state (see below and SI, section~\ref{theory:
Ising_SOC_perturbation} for discussion of interactions) and in-plane depairing
effects. As proof of concept, we consider a simple model that incorporates two
depairing mechanisms: Rashba SOC (which favours in-plane spin orientation) and
orbital in-plane magnetic field effects---both of which compete with the Ising
SOC and suppress the PVR (see SI, section \ref{theory: BCS_meanfield}). While
no direct signatures of Rashba SOC are observed in our sample, quantum
oscillations at low $|D|$ place an upper bound on the Rashba SOC parameter
$\lambda_R \lesssim 4$~meV (see Methods and \prettyref{exfig:QHE_ising}),
consistent with previous studies reporting $\lambda_R$ ranging from $1$ to $15$
meV\cite{gmitraProximityEffectsBilayer2017, wangOriginMagnitudeDesigner2016,yangStrongElectronholeSymmetric2017,
liTwistangleDependenceProximity2019, amannCounterintuitiveGateDependence2022}.
The solution of a self-consistent superconducting gap equation for this model
can capture the observed PVR evolution (\prettyref{fig:Fig4}c inset and
\prettyref{exfig:theory_figure_PLV}), e.g., if the effective Rashba spin
splitting increases with hole density in the $\mathrm{FP}(2,2)_{+}$ phase. Such
an increase is expected if superconductivity arises from minority Fermi pockets
that grow with hole doping (see SI, sections~\ref{theory: BCS_meanfield} and
\ref{theory: PVR_fit} for details and comparison to experimental data, as well
as a discussion of orbital in-plane field effects).
  
The extended phase space of superconductivity in BLG-WSe$_2$ clearly contrasts
observations in hBN-encapsulated crystalline bilayer and trilayer
graphene\cite{zhouIsospinMagnetismSpinpolarized2022,
zhouSuperconductivityRhombohedralTrilayer2021}, where superconductivity is
observed only within a narrow density range around the symmetry-broken phase
boundaries. Moreover, the coincidence of the doping range exhibiting
superconductivity with the $\mathrm{FP}(2,2)_{+}$ phase
(\prettyref{fig:Fig3}c,e) at $D>0$ strongly hints that $(i)$ superconductivity
descends from the latter broken-symmetry parent state and $(ii)$ SOC plays a
key role in selecting a symmetry-breaking order conducive to pairing.
Figure~\ref{fig:Fig4}f depicts a phenomenologically motivated scenario wherein
multiple nearly degenerate broken-symmetry orders compete. If the
$\mathrm{FP}(2,2)_{+}$ phase is, e.g., valley polarized in the absence of SOC,
then broken inversion and time-reversal symmetries would heavily disfavour
pairing---consistent with the absence of superconductivity in BLG-WSe$_2$ at
$D<0$ and hBN-encapsulated BLG at zero magnetic
field\cite{zhouIsospinMagnetismSpinpolarized2022}. Turning on Ising SOC could
then tip the balance in favour of orders that facilitate Cooper pairing by
restoring resonance between opposite-momentum states along the Fermi surfaces.
For instance, a spin-valley polarized state in which interactions enhance the
bare Ising SOC strength to produce the observed large and small Fermi surfaces
could naturally host Ising superconductivity; such a state would, however,
exhibit much stronger Pauli-limit violation than is observed and can thus be
ruled out. Alternatively, we suggest that Ising SOC can promote intervalley
coherent (IVC) order that is \emph{also} amenable to pairing while maintaining
compatibility with observed Pauli-limit violation trends (see SI,
section~\ref{theory: Ising_SOC_perturbation}). Field-induced spin-polarized
superconductivity in hBN-encapsulated BLG may analogously arise if the Zeeman
energy destabilizes valley polarization near the broken-symmetry phase
boundary.
  
The nature of superconductivity in graphene-based systems---both moir\'e and
crystalline\cite{dongSuperconductivityVicinityIsospinpolarized2021,
ghazaryanUnconventionalSuperconductivitySystems2021,
qinFunctionalRenormalizationGroup2022,
youKohnLuttingerSuperconductivityIntervalley2022,
ceaSuperconductivityRepulsiveInteractions2022,
chouAcousticphononmediatedSuperconductivityRhombohedral2021,
chouAcousticphononmediatedSuperconductivityBernal2022}---presents an ongoing
puzzle. Our work demonstrates that induced SOC can enhance $T_c$ in BLG by an
order of magnitude, while also stabilizing superconductivity over a much wider
parameter space that crucially includes zero magnetic field. This behaviour is
reminiscent of earlier works in twisted bilayer graphene coupled to WSe$_2$
where superconductivity persisted far away from the magic
angle\cite{aroraSuperconductivityMetallicTwisted2020}. Moreover, an enticing
general similarity between BLG-WSe$_2$ and moir\' e graphene
superlattices\cite{caoUnconventionalSuperconductivityMagicangle2018,
parkTunableStronglyCoupled2021,haoElectricFieldTunable2021,
zhangAscendanceSuperconductivityMagicAngle2021,
parkMagicAngleMultilayerGraphene2021} can be noticed, as in both systems
superconductivity appears intimately connected to the symmetry-broken state in
which two out of four spin-valley flavours are predominately populated. In this
context, our results provide guidance for future efforts aiming to address the
origin of apparent striking distinctions between different superconducting
phases in graphene systems. Finally, induced SOC parameters depend on the
relative orientation of WSe$_2$ (or other TMDs) and
graphene\cite{liTwistangleDependenceProximity2019}, and are thus in principle
tunable---providing a rich landscape for further exploring the interplay
between spin-orbit effects, correlated phases, and superconductivity in
ultra-clean crystalline graphene multilayers.


\noindent {\bf References:}


\clearpage
\clearpage

\clearpage

\clearpage

\section*{Methods}

\textbf{Device fabrication:} Both devices have a dual-graphite gate structure
with graphite electrodes, and were assembled as follows: First, a thin hBN
flake ($10-30$~nm) is picked up using a propylene carbonate (PC) film
previously placed on a polydimethylsiloxane (PDMS) stamp. Then, the hBN flake
is used to pick up crystals in the sequence of graphite top gate, top hBN
dielectric, an exfoliated monolayer of WSe$_2$ (commercial source, HQ
graphene), Bernal bilayer graphene, graphite electrodes, bottom hBN dielectric,
and graphite bottom gate. Care was taken to approach and pick up each flake
slowly. In the last step, the whole stack is dropped onto a Si/SiO$_2$
substrate at $150\degree$C while the PC is released at $180\degree$C. The PC is
then cleaned off with N-Methyl-2-Pyrrolidinone (NMP). The final geometry is
defined by dry etching with a CHF$_3$/O$_2$ plasma and deposition of ohmic edge
contacts (Ti/Au, 5 nm/100 nm); see \prettyref{exfig:reproducibility}.

\textbf{Measurements:} All measurements were performed in a dilution
refrigerator (Oxford Triton) with a base temperature of $\sim30$~mK, using
standard low-frequency lock-in amplifier techniques. Unless otherwise
specified, measurements are taken at the base temperature. Frequencies of the
lock-in amplifiers (Stanford Research, models 865a) were kept in the range of
$7-40$~Hz in order to reduce the electronic noise and measure the device's DC
properties. The AC excitation was kept $<5$~nA (most measurements were taken at
$0.5-1$~nA to preserve the linearity of the system and avoid disturbing the
fragile states at low temperatures). Each of the DC fridge lines pass through
cold filters, including 4 Pi filters that filter out a range from $\sim80$~MHz
to $>10$~GHz, as well as a two-pole RC low-pass filter.

\textbf{Reproducibility of zero-magnetic-field superconductivity:}
\prettyref{exfig:reproducibility}b,c shows optical images of BLG-WSe$_2$
devices. We use a dual-graphite gate structure to minimize charge
disorder\cite{zibrovRobustFractionalQuantum2017}. Superconductivity and
symmetry-breaking features are exactly the same between different contacts
in one device (\prettyref{exfig:reproducibility}d,e), thanks to the
exceptionally high quality of crystalline graphene. Contacts 1-3 of the first
device D1 were used for the measurements in the main text. The second device D2
reproduces the zero-magnetic-field superconductivity with similar doping ranges
(\prettyref{exfig:reproducibility}f,g). Slight differences between the two
devices could originate from different SOC
strengths\cite{liTwistangleDependenceProximity2019} induced by WSe$_2$. We
fabricated four BLG-WSe$_2$ devices in total, and two of them show
zero-magnetic-field superconductivity. The devices that do not exhibit
superconductivity have different overall $n$--$D$ phase diagrams (see
\prettyref{exfig:nonSCdevice}), suggesting that the symmetry-broken ground
states selected by different SOC strengths are distinct and not always
conducive to pairing.

\begin{figure}[p]
    \includegraphics[width=16cm]{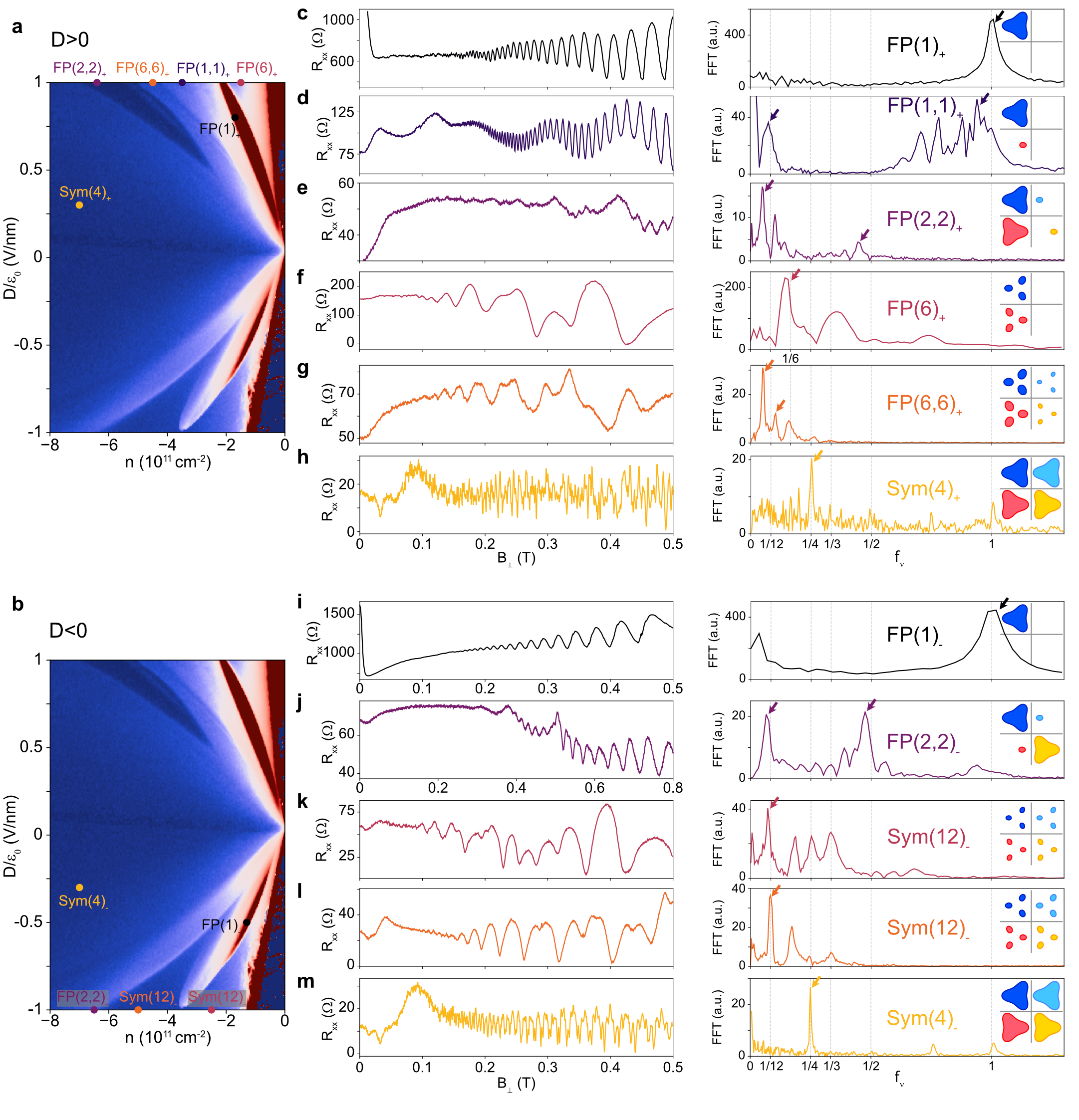}
    \centering
    \caption{{\bf Quantum oscillations at selected gate points.} 
    {\bf a},{\bf b}, $n$--$D$ phase diagram with the coloured dots indicating the positions at which the quantum oscillations (left panels in {\bf c}--{\bf m}) are taken.
    {\bf c}--{\bf m}, Left panels show the quantum oscillations at the coloured dots. Right panels show the normalized Fourier transform of the corresponding $R_{xx}(1/B_\perp)$ data.}
\label{exfig:QO_cuts}
\end{figure}

\textbf{Identifying different spin-valley flavour-polarized phases:}
BLG-WSe$_{2}$ realizes rather complex spin-valley flavour-polarized phases for
both positive and negative $D$ fields (\prettyref{exfig:QO_cuts},
\ref{exfig:negativeDfield} and \ref{exfig:Fan and FFT}). We argue
that the $D<0$ phase diagram is similar to that of hBN-encapsulated BLG, while
the $D>0$ phase diagram has essential differences associated with the interplay
between SOC and strong correlations.
  
For $D<0$, at low $|D|$ and high $|n|$, fast Fourier transform (FFT) shows a
prominent peak at $f_{\nu} =1/4$ corresponding to a flavour-symmetric phase
that preserves the four-fold spin-valley degeneracy ($\mathrm{Sym}(4)_{-}$;
\prettyref{exfig:QO_cuts}m). The spin-valley symmetry still holds for high
$|D|$ and low $|n|$, but smaller Fermi pockets are produced by trigonal warping
within each flavour, and therefore the system is flavour-symmetric with
$f_{\nu} =1/12$ ($\mathrm{Sym}(12)_{-}$; \prettyref{exfig:QO_cuts}k). As
mentioned in the main text, the diagonal largest-resistance region is a single
spin-valley flavour-polarized phase ($\mathrm{FP}(1)_{-}$;
\prettyref{exfig:QO_cuts}i) that peaks at $f_{\nu} =1$. The remaining
flavour-polarized phases have multiple Fermi pockets with distinct Fermi
surface areas. At slightly higher $|n|$ adjacent to $\mathrm{FP}(1)_{-}$, the FFT
in the region exhibits peaks around $f_{\nu}^{(1)} <1$ and $f_{\nu}^{(2)}
<1/12$. This is a flavour-polarized phase with one majority flavour and one (or
more) small Fermi pocket ($\mathrm{FP}(1,1)_{-}$). At the region we observed
the nonlinear resistive phase, the Fermi surface has two frequency peaks near
$f_{\nu}^{(1)} < 1/2$ and $f_{\nu}^{(2)} < 1/12$ such that
$f_{\nu}^{(1)}+f_{\nu}^{(2)} = 1/2$, and corresponds to a flavour-polarized
phase with two majority ($f_{\nu}^{(1)} < 1/2$) and two minority
($f_{\nu}^{(2)} < 1/12$) flavours ($\mathrm{FP}(2,2)_{-}$;
\prettyref{exfig:QO_cuts}j). At lower $|n|$ next to $\mathrm{FP}(2,2)_{-}$, a
flavour-symmetric phase emerges with $12$ trigonally warped pockets
($\mathrm{Sym}(12)_{-}$; \prettyref{exfig:QO_cuts}l).
 
At $D>0$, by contrast, the spin degeneracy in each valley is explicitly lifted
by Ising SOC (\prettyref{fig:Fig1}b). At high $D$ and low $|n|$, instead of
showing frequency at $f_{\nu} =1/12$, quantum oscillations at the band edge
exhibit a peak around $f_{\nu} =1/6$ (\prettyref{exfig:QO_cuts}f), and this is
consistent with Ising-induced spin splitting such that holes are from small
trigonally warped Fermi pockets of single spin species in each valley
($\mathrm{FP}(6)_{+}$). The region next to $\mathrm{FP}(2,2)_{+}$ also shows
different frequencies ($\mathrm{FP}(6,6)_{+}$; \prettyref{exfig:QO_cuts}g):
this can be attributed to Ising-induced band splitting with one spin more filled
($f_{\nu}^{(1)} >1/12$) and another spin less filled ($f_{\nu}^{(2)} <1/12$) in
each valley. Flavour-polarized phases, such as $\mathrm{FP}(1)_{+}$,
$\mathrm{FP}(1,1)_{+}$, and $\mathrm{FP}(2,2)_{+}$
(\prettyref{exfig:QO_cuts}c,d,e), are overall not changed much in terms of FFT
frequencies, though spin-valley configurations are most likely different from
the $D<0$ cases.

\textbf{Similarity to hBN-encapsulated BLG at $D <0$:} As discussed in the
previous section, at $D <0$ the symmetry-broken phases resemble those observed
in hBN-encapsulated BLG. In $\mathrm{FP}(2,2)_{-}$, we observed the resistive
phase showing nonlinear critical current behaviour (\prettyref{fig:Fig1}f) at
zero magnetic field. The similarity is also supported by fan diagrams and FFT
(\prettyref{exfig:QO_cuts} and \ref{exfig:negativeDfield}) since
flavour-symmetric and flavour-polarized states observed in hBN-encapsulated BLG
are well reproduced at $D <0$. However, we did not observe superconductivity
with finite in-plane magnetic field at $D<0$. The absence of field-induced
superconductivity in this regime may reflect %
of slightly higher electron temperature ($\sim30$~mK) and small in-plane-field
misalignment. We thus can not rule out the onset of superconductivity upon more
careful characterization.  Alternatively, Rashba SOC (which contrary to Ising SOC need
not be suppressed at $D<0$) is expected to compete against spin polarization
favoured by an in-plane field, thus potentially precluding field-induced
superconductivity.

\begin{figure}[p]
    \includegraphics[width=16cm]{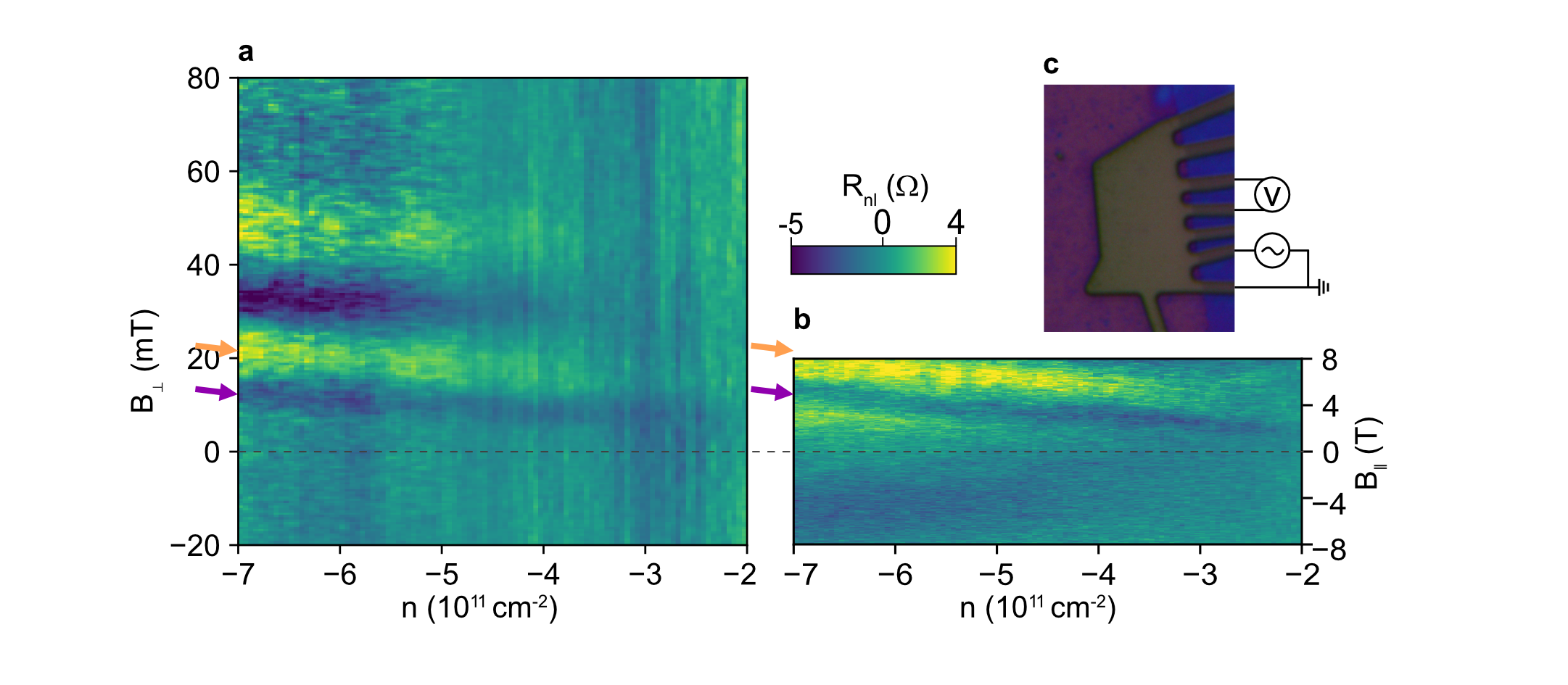}
    \centering
    \caption{{\bf Transverse magnetic focusing with out-of-plane and in-plane magnetic field.} {\bf a}, Non-local resistance $R_{nl}$ measured as a function of $n$ and $B_{\perp}$ at $D/\epsilon_0= 0.6$~V/nm with the configuration shown in {\bf c}. {\bf b}, Non-local resistance $R_{nl}$ measured as a function $n$ and $B_{\parallel}$ at $D/\epsilon_0= -0.2$~V/nm. Transverse magnetic focusing with an in-plane field is due to imperfect sample alignment. Therefore, we can estimate the field misalignment angle by comparing {\bf a} and {\bf b} (see Methods for further discussion).}
\label{exfig:magnetic focusing}
\end{figure}

\begin{figure}[p]
    \includegraphics[width=0.78\linewidth]{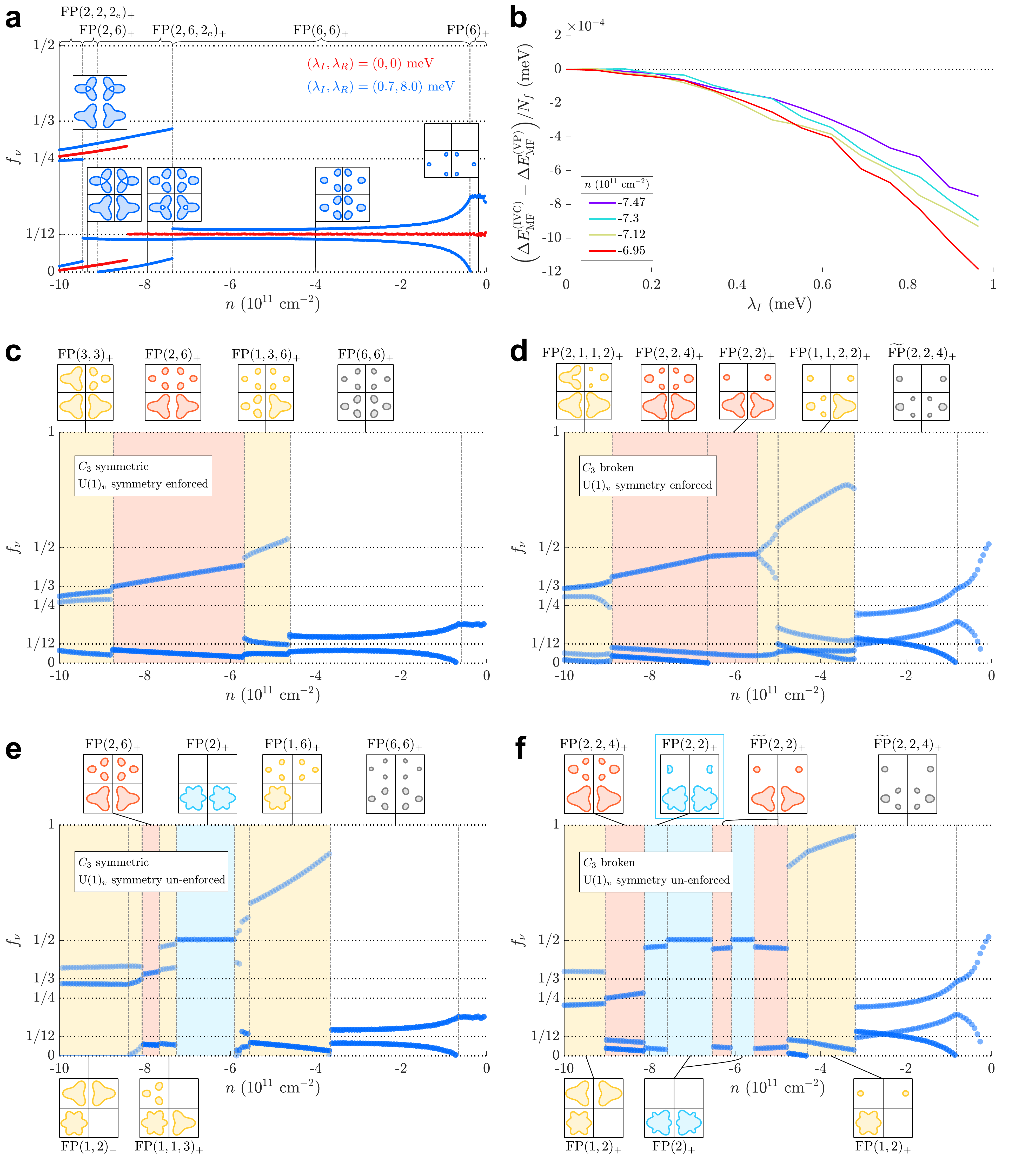}
    \centering
    \caption{{\bf Polarized phases.}
    {\bf a},{\bf c}-{\bf f}, Expected normalized quantum oscillation frequencies as a function of density without interactions ({\bf a}), allowing for $\mathrm{U}(1)_v$ unbroken states with $C_3$ preserved ({\bf c}), allowing for $\mathrm{U}(1)_v$ unbroken states with $C_3$ broken ({\bf d}),  allowing for IVC ordered states with $C_3$ preserved ({\bf e}), and allowing IVC ordered states with $C_3$ broken ({\bf f}). 
    Regions with red (blue) backgrounds correspond to singly polarized states that preserve  (spontaneously break) the valley symmetry (even when the initial conditions were chosen to allow IVC order to develop, as in {\bf e} and {\bf f}).
    Regions coloured yellow are multiply polarized (no distinction is made between those with and without IVC order).
    Insets show the Fermi surfaces corresponding to a select set of fillings.
    The $D$ field is set to $D=1$~V/nm in all plots.
    All simulations include SOC ($\lambda_I=0.7$~meV and $\lambda_R=3$~meV) except for the red curve in {\bf a}. 
    {{\bf b},} Plot of the difference between the change in ground state energy induced by Ising SOC for an IVC state and the change in ground state energy induced by Ising SOC for a VP state. The energy is normalized by the number of carriers $N_f$.
    The negative values obtained imply that the addition of Ising SOC to a VP state increases its energy more than the addition of Ising SOC increases the ground state energy of an IVC ground state.
    }
\label{exfig:theory_figure_cascade}
\end{figure}

\begin{figure}[p]
    \includegraphics[width=16cm]{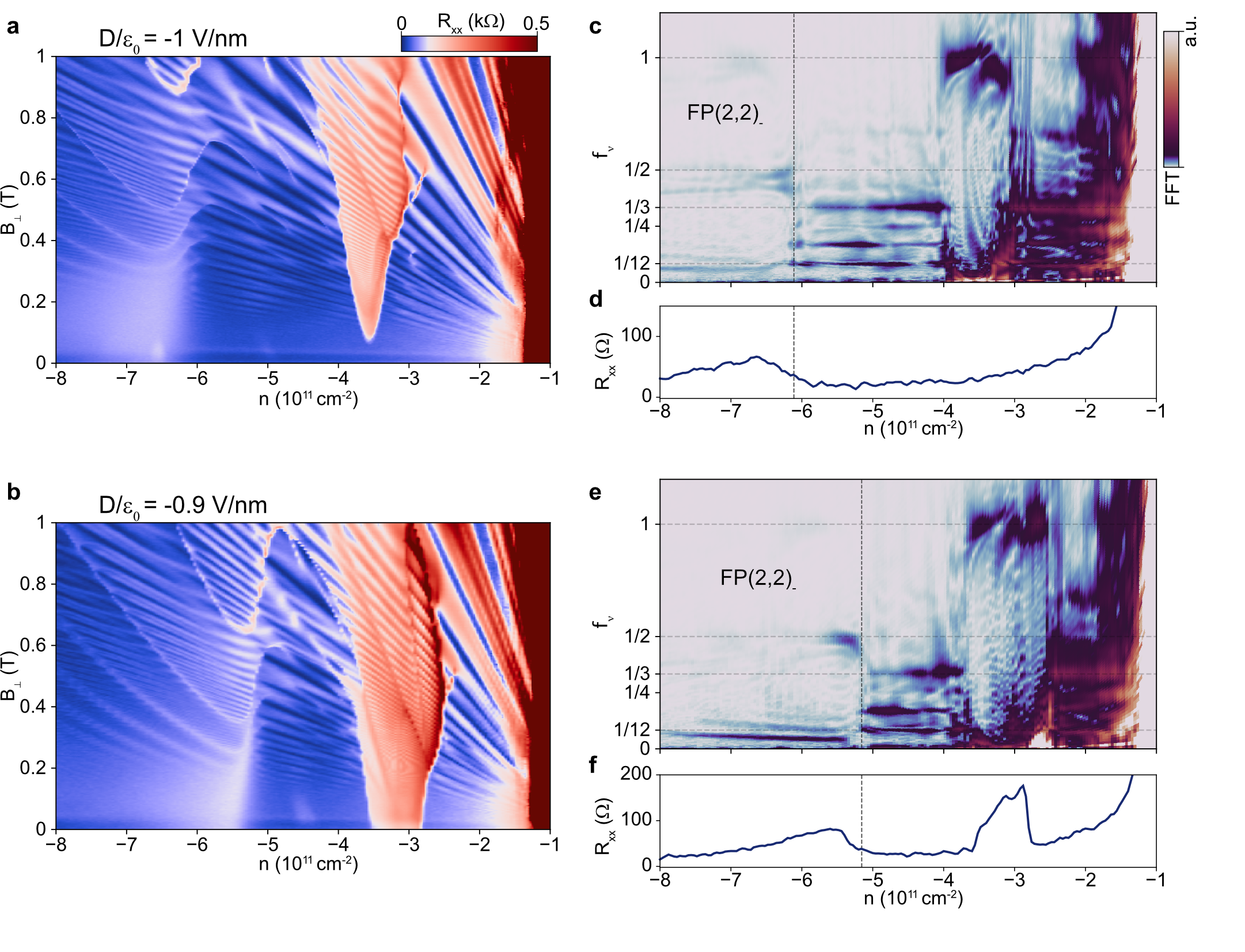}
    \centering
    \caption{{\bf Fan diagrams at $D<0$.} {\bf a},{\bf b}, $R_{xx}$ versus $B_\perp$ and doping density $n$ for  $D/\epsilon_0= -1$~V/nm ({\bf a}) and $-0.9$~V/nm ({\bf b}), respectively. {\bf c},{\bf e}, Fourier transform of $R_{xx}(1/B_\perp)$ versus $n$ and $f_\nu$ for $D/\epsilon_0= -1$~V/nm and $-0.9$~V/nm, respectively. $R_{xx}$ data within $0.05$~T~$<B_\perp<0.6$~T are used for converting. The corresponding $R_{xx}$ data at zero magnetic field are shown in {\bf d} and {\bf f}.}
\label{exfig:negativeDfield}
\end{figure}

\begin{figure}[p]
    \includegraphics[width=16cm]{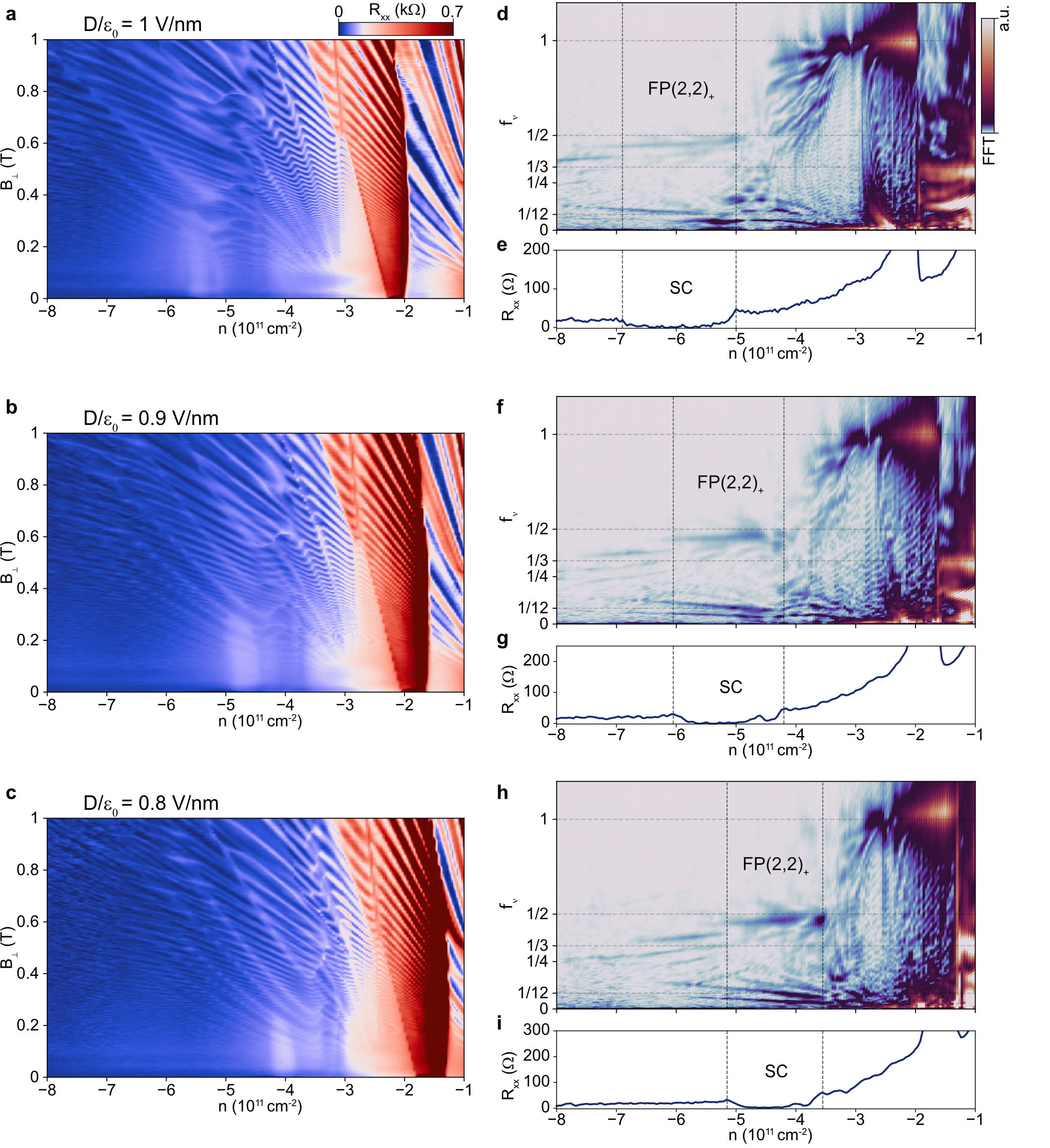}
    \centering
    \caption{{\bf Zero-magnetic-field superconductivity and $\mathrm{FP}(2,2)_{+}$ phase at $D>0$.} {\bf a}--{\bf c}, $R_{xx}$ versus $B_\perp$ and doping density $n$ for  $D/\epsilon_0= 1$~V/nm ({\bf a}), $0.9$~V/nm ({\bf b}), and $0.8$~V/nm ({\bf c}), respectively. {\bf d},{\bf f},{\bf h}, Fourier transform of $R_{xx}(1/B_\perp)$ versus $n$ and $f_\nu$ for $D/\epsilon_0= 1$~V/nm, $0.9$~V/nm, and $0.8$~V/nm, respectively. $R_{xx}$ data within $0.05$~T~$<B_\perp<0.6$~T are used for converting. The corresponding $R_{xx}$ data at zero magnetic field are shown below ({\bf e},{\bf g},{\bf i}). We see a good match between the doping range exhibiting superconductivity and the $\mathrm{FP}(2,2)_{+}$ phase region, regardless of $D$ fields.}
\label{exfig:Fan and FFT}
\end{figure}

\textbf{Transverse magnetic focusing with out-of-plane magnetic field:} The mean free path
$\ell_{mf}$ of BLG-WSe$_2$ is around $10$~$\mu$m. \prettyref{exfig:magnetic
focusing}a shows non-local resistance $R_{nl}$ as a function of $n$ and
$B_\perp$ for $D/\epsilon_0= 0.6$~V/nm measured with the configuration shown in
\prettyref{exfig:magnetic focusing}c. Data at density $n = -7\times 10^{11}
~\text{cm}^{-2}$ show a pronounced feature around $B_{\perp} \approx 20$~mT,
which suggests a transverse magnetic
focusing\cite{taychatanapatElectricallyTunableTransverse2013} that is
comparable with the electrodes separation of 5 $\mu$m, and translates to a mean
free path $\ell_{mf} \gtrsim \pi L/2 \approx 7.9~\mu$m. The magnetic focusing
feature appears over wide density ranges, including the density ($\sim-3\times
10^{11} ~\text{cm}^{-2}$) where superconductivity is observed at this $D$
field.
 
\textbf{Sample alignment with in-plane magnetic field:} In-plane-field
measurements were performed by mounting the sample vertically with a homemade
frame. It is inevitable to introduce a small $B_{\perp}$ component when the
$B_{\parallel}$ field is applied due to the imperfect vertical sample
alignment. Transverse magnetic focusing in $B_{\parallel}$ is a reliable
measurement for the angle misalignment since the cyclotron orbits only couple
to the $B_{\perp}$ component. \prettyref{exfig:magnetic focusing}b shows
$R_{nl}$ as a function of $n$ and $B_\parallel$. The $B_{\parallel}$ plot
qualitatively matches the $B_{\perp}$ plot (\prettyref{exfig:magnetic
focusing}a) except the scaling of the $B$ field axis. The $R_{nl}$ peak feature
that appears at $B_{\perp} \approx 20$~mT roughly matches the same feature in
in-plane field at $B_{\parallel} \approx 7$~T. This suggests an in-plane-field
misalignment angle $\theta_{mis} \approx
\text{tan}^{-1}(20~\text{mT}/7~\text{T}) \approx 0.16 \degree$.
 
\begin{figure}[p]
    \includegraphics[width=16cm]{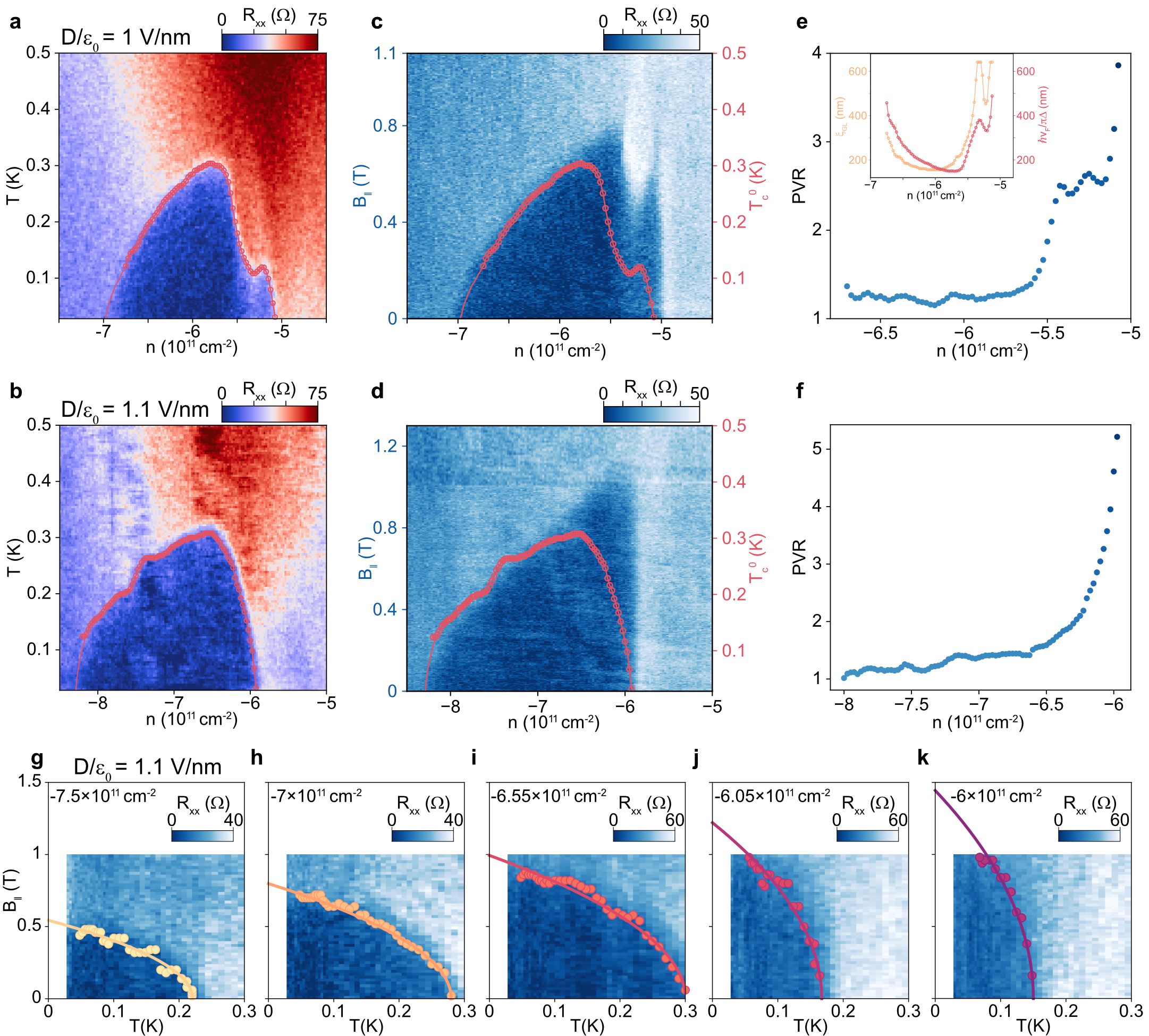}
    \centering
    \caption{{\bf In-plane magnetic field dependence of superconductivity.} 
    {\bf a},{\bf b}, $R_{xx}$ versus doping density $n$ and temperature for the superconducting domes at $D/\epsilon_0 = 1~\text{V/nm}$ ({\bf a}) and  $1.1~\text{V/nm}$ ({\bf b}), respectively. 
    {\bf c},{\bf d}, $R_{xx}$ versus $n$ and $B_{\parallel}$ for the superconducting domes at $D/\epsilon_0 = 1~\text{V/nm}$ ({\bf c}) and  $1.1~\text{V/nm}$ ({\bf d}), respectively. 
    Red dots indicate the critical temperature at zero magnetic field. 
    {\bf e},{\bf f}, Pauli violation ratio (PVR) calculated from $B_{c\parallel}^{T \approx 30~\text{mK}}/B_p$ as a function of doping density $n$. 
    Both curves feature strong Pauli-limit violation at low $|n|$.
    Inset of {\bf e} shows coherence length $\xi_\mathrm{GL}  = \sqrt{\Phi_0/(2\pi B_{c\perp})}$ and $\hbar v_F/\pi \Delta$ versus $n$ at $D/\epsilon_0 = 1~\text{V/nm}$. $\hbar v_F/\pi \Delta$ is estimated with a weak-coupling assumption: $\Delta\approx1.76k_B T_c$ and $v_F = \hbar k_f/m^\ast$ ($k_f = \sqrt{2\pi f_{\nu}|n|}$ with $f_{\nu}$ being the normalized frequency of minority Fermi pockets, $m^\ast \sim 0.15m_e$).
    {\bf g}--{\bf k}, $R_{xx}$ versus temperature and $B_{\parallel}$ at different densities for $D/\epsilon_0 = 1.1~\text{V/nm}$. 
    In-plane critical fields $B_{c\parallel}$ (the field at which $R_{xx}$ is $50\%$ of the normal state resistance) are marked by dots, and all the data are well-fit by the phenomenological relation. 
    We see a clear evolution of the PVR as a function of doping.}
\label{exfig:inplanefield}
\end{figure}

Such angle misalignment results in an underestimation of in-plane critical
field at regions where $B_{c\perp}$ is small, i.e., near the phase boundaries.
\prettyref{exfig:inplanefield}c shows $R_{xx}$ versus $n$ and $B_{\parallel}$
measured at $D/\epsilon_0= 1$~V/nm. At the phase boundary between
$\mathrm{FP}(6,6)_{+}$ and $\mathrm{FP}(2,2)_{+}$ ($n\approx-5\times 10^{11}
~\text{cm}^{-2}$), superconductivity disappears around $B_{\parallel} = 0.7$~T,
which suggests an out-of-plane field component $B_{\perp} = 0.7~\text{T} \times
\text{tan}(0.16 \degree) \approx 2$~mT. The $B_{\perp} = 2$~mT component
roughly matches $B_{c\perp}$ at the same density (see \prettyref{fig:Fig2}f).
Therefore, we conclude that the Pauli violation ratio around the phase
boundaries in \prettyref{fig:Fig4}c only serves as a lower limit since
$B_{c\perp}$ is rather low at the relevant densities and hence $B_{\perp}$ is a
main driver for superconductivity suppression at those regions. By contrast, at
the density range where $B_{c\parallel}$ is roughly consistent with the Pauli
limit (higher $|n|$), superconductivity shows much higher $B_{c\perp}$ (see
\prettyref{fig:Fig2}f). The suppression of superconductivity is then mainly
caused by $B_{\parallel}$ at higher $|n|$.

\begin{figure}[p]
    \includegraphics[width=16cm]{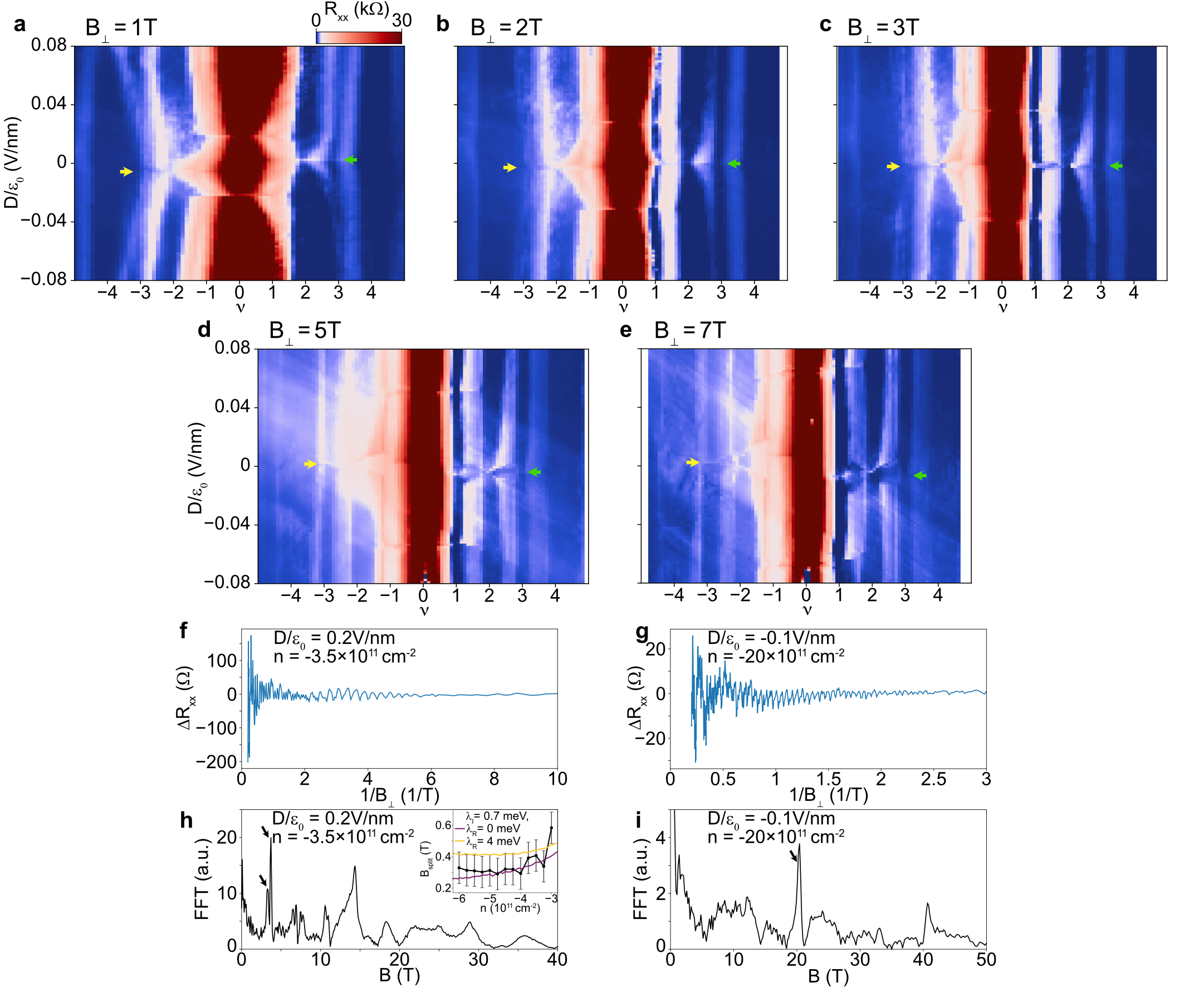}
    \centering
    \caption{{\bf Estimating different SOC strengths through quantum Hall effect and quantum oscillations.} 
    {\bf a}--{\bf e}, $R_{xx}$ versus $\nu = 2\pi \ell_{B}^{2}n$ ($\ell_B$ is the Landau magnetic length) and $D$ field at $B_\perp = 1$~T, $2$~T, $3$~T, $5$~T, and $7$~T, respectively. Arrows mark the transition of $|\nu| = 3$ quantum Hall states with $D$ field. Because the Ising SOC is oriented out of plane, an out-of-plane Zeeman splitting will cancel it when $2E_Z = 2g\mu_B B_\perp^* = \lambda_I$ ($B_\perp^*$ is the magnetic field at which yellow and green arrows are at the same $D$ field; $B_\perp^* \approx3$~T here). {\bf f},{\bf g}, $\Delta R_{xx}$ versus $1/B_{\perp}$ (measured up to $B_{\perp}=5$~T) at $D/\epsilon_0= 0.2$~V/nm, $n = -3.5\times 10^{11} ~\text{cm}^{-2}$ ({\bf f}) and $D/\epsilon_0= -0.1$~V/nm, $n = -20\times 10^{11} ~\text{cm}^{-2}$ ({\bf g}), respectively. The corresponding FFT data are shown in {\bf h} and {\bf i}. Inset of {\bf h} shows the FFT splitting $B_\text{split}$ (marked by black arrows in the main panel) versus doping density $n$ measured at $D/\epsilon_0= 0.2$~V/nm. Coloured lines show the FFT splitting predicted from band structure calculations for the same $D$ field, using Ising SOC $\lambda_I = 0.7$~meV with Rashba SOC $\lambda_R = 0$~meV (purple line) and $\lambda_R = 4$~meV (yellow line).}
\label{exfig:QHE_ising}
\end{figure}

\begin{figure}[p]
    \includegraphics[width=15cm]{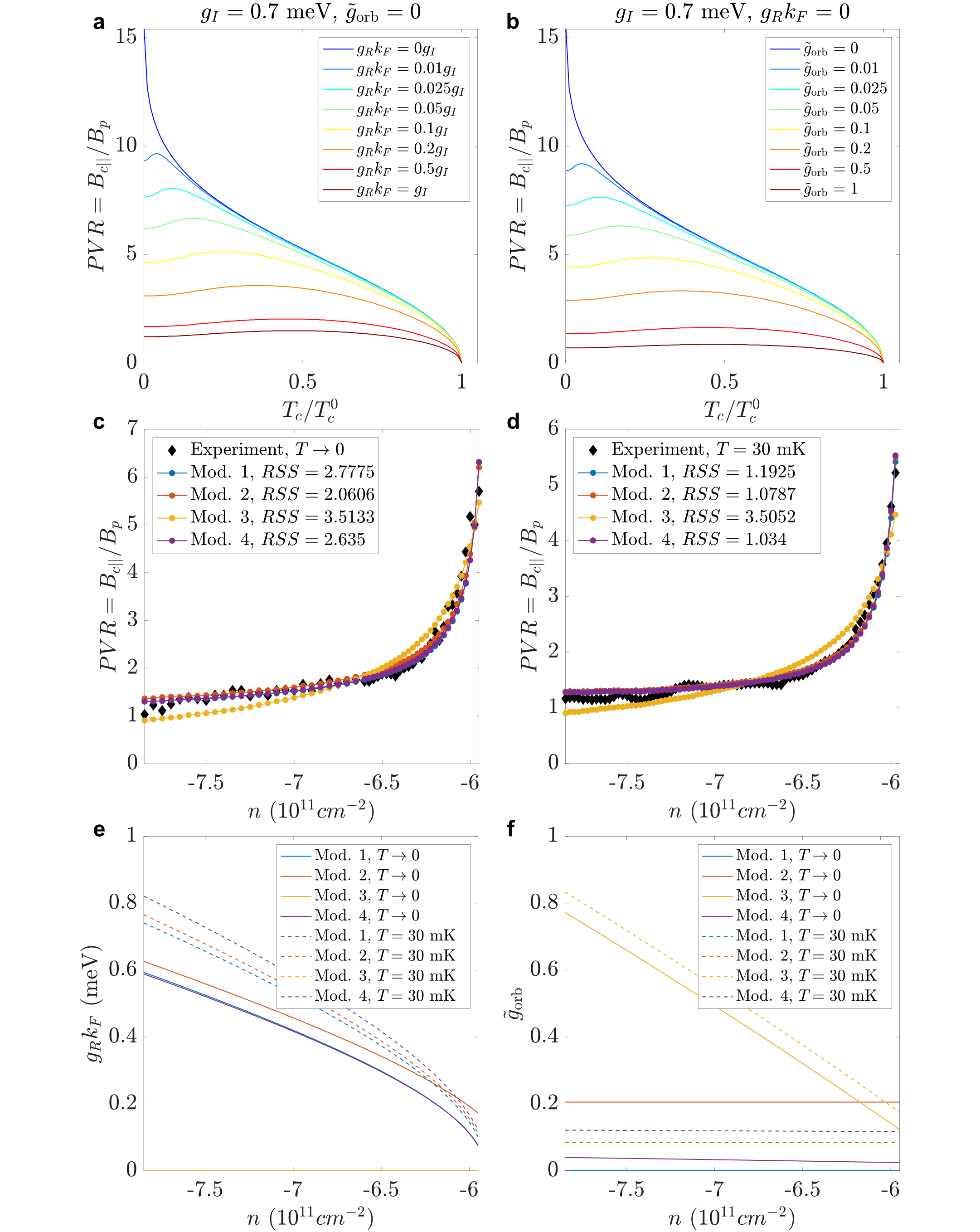}
    \centering
    \caption{ {\bf Depairing model for doping-dependent Pauli violation ratio.} {\bf a},{\bf b}, Pauli violation ratio expected in a system with: Ising $g_I$ and Rashba $g_R k_F$ coupling ({\bf a}), Ising $g_I$ and orbital $\tilde{g}_{\rm orb}$ coupling ({\bf b}). Note that $\tilde{g}_{\rm orb}$ is a dimensionless quantity: the corresponding orbital energy scale is $\tilde{g}_{\rm orb} \mu_B B$. {\bf c},{\bf d}, Fitting the model described in SI, section \ref{theory: BCS_meanfield} to the experimental data in \prettyref{fig:Fig4}c ({\bf c}) and \prettyref{exfig:inplanefield}f ({\bf d}). {\bf e},{\bf f}, Evolution of the extracted parameters $g_R k_F$ and $\tilde{g}_{\rm orb}$ as a function of hole density $n$, for the four models that are used in the fitting procedure (see SI, section \ref{theory: PVR_fit}); $g_I = 0.7$ meV in all the plots.}
\label{exfig:theory_figure_PLV}
\end{figure}

\begin{figure}[p]
    \includegraphics[width=16cm]{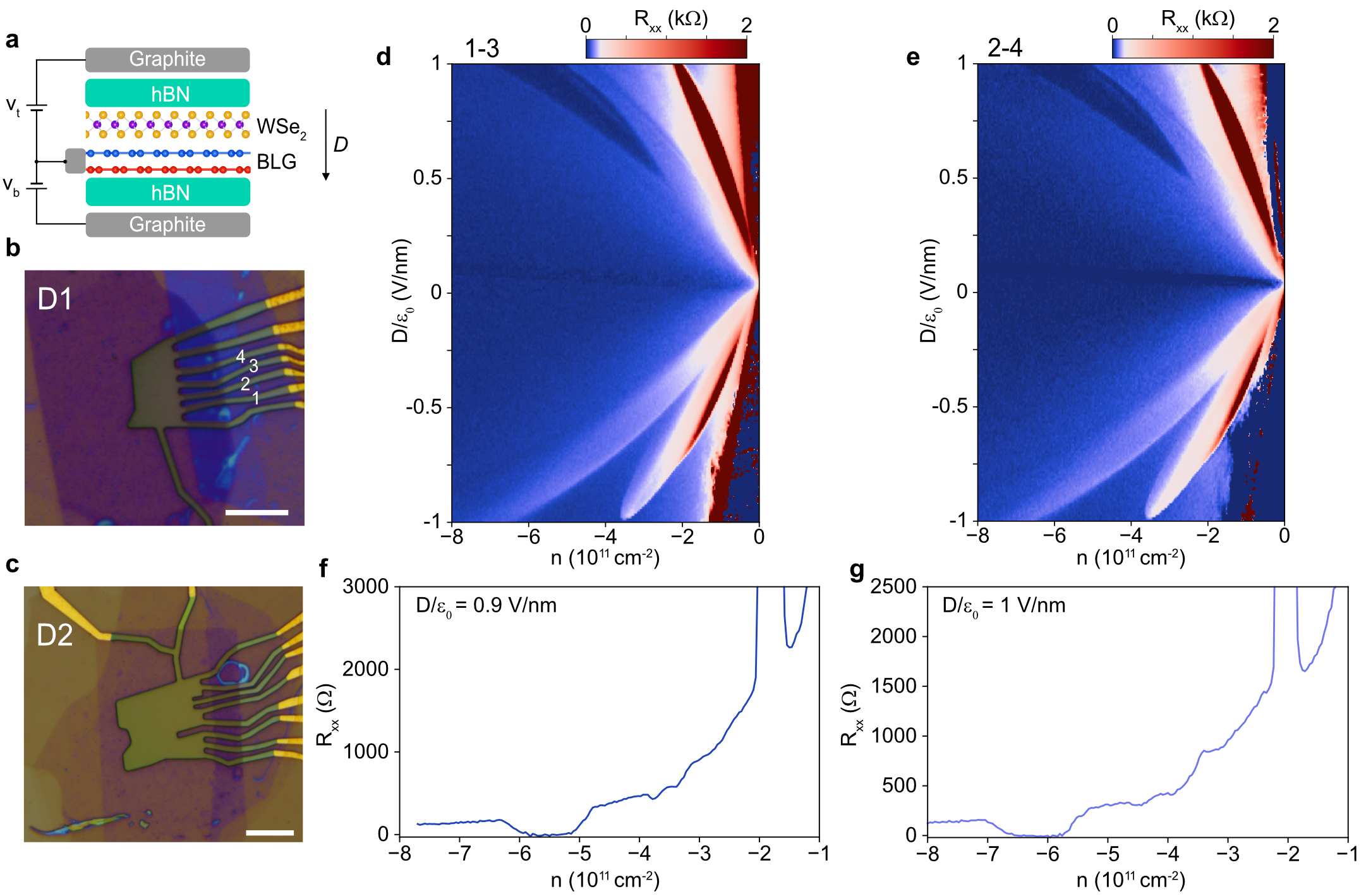}
    \centering
    \caption{{\bf Reproducibility of zero-magnetic-field superconductivity in BLG-WSe$\mathbf{_{2}}$.} {\bf a}, Schematic of a dual-gated device. Doping density $n$ and $D$ field are controlled by tuning top and bottom gate voltage $v_{t}$ and $v_{b}$.  {\bf b},{\bf c}, Optical images of the investigated devices. The scale bar in each panel corresponds to $10~\mu$m. {\bf d},{\bf e}, $R_{xx}$ versus doping density $n$ and $D$ field measured from the first device D1 between contacts 1-3 ({\bf d}) and 2-4 ({\bf e}), respectively. Contacts 1-3 were used for the measurements in the main text. {\bf f},{\bf g}, $R_{xx}$ versus doping density $n$  measured from the second device D2 at $D/\epsilon_0= 0.9$~V/nm ({\bf f}) and $1$~V/nm ({\bf g}), respectively.}
\label{exfig:reproducibility}
\end{figure}

\begin{figure}[p]
    \includegraphics[width=16cm]{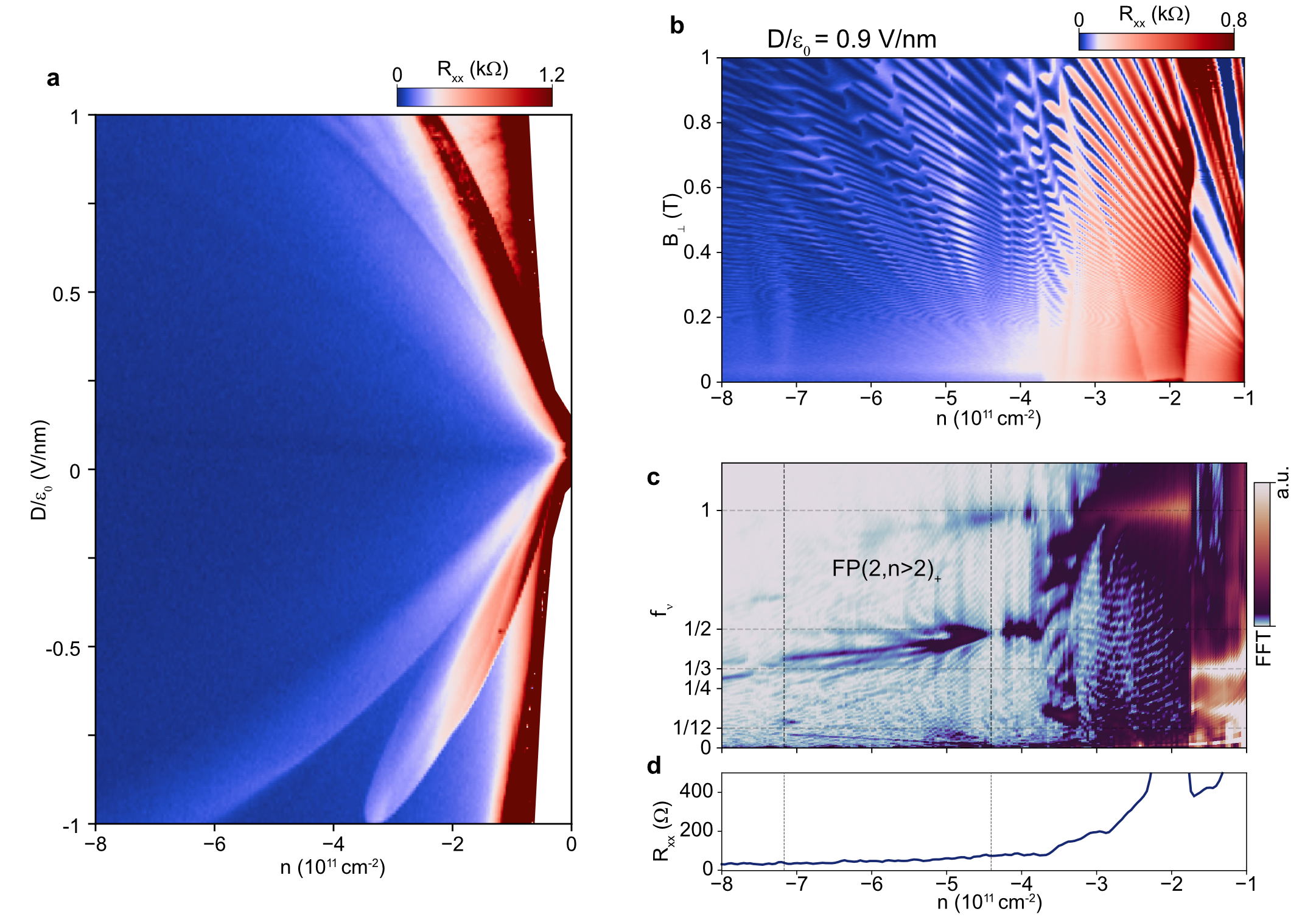}
    \centering
    \caption{{\bf Phase diagram of a BLG-WSe$\mathbf{_2}$ device without zero-magnetic-field superconductivity.} {\bf a}, $R_{xx}$ versus doping density $n$ and displacement field $D$ for a BLG-WSe$_2$ device without zero-magnetic-field superconductivity. {\bf b}, $R_{xx}$ versus $B_\perp$ and doping $n$ for $D/\epsilon_0= 0.9$~V/nm. {\bf c}, Fourier transform of $R_{xx}(1/B_\perp)$ versus $n$ and $f_\nu$ for $D/\epsilon_0= 0.9$~V/nm. The corresponding $R_{xx}$ data at zero magnetic field is shown in {\bf d}. The non-superconducting device possesses a flavour-polarized phase with two majority and multiple ($n>2$) minority Fermi pockets (denoted as $\mathrm{FP}(2,n>2)_{+}$). The observation thus further substantiates $\mathrm{FP}(2,2)_{+}$ as a parent state for zero-magnetic-field superconductivity.}
\label{exfig:nonSCdevice}
\end{figure}

\textbf{Ising SOC:} In the main text, we show that the asymmetric $n$--$D$
phase diagram provides strong evidence of Ising SOC. Quantum oscillations of
the non-interacting phases at $D>0$ further support the existence of Ising SOC
(see \prettyref{exfig:QO_cuts} and \ref{exfig:QHE_ising}). To quantify
WSe$_{2}$-induced Ising SOC, we probe the octet zeroth Landau level (LL) in
BLG, since few-meV-scale Ising SOC can rearrange the energies of these states.
Note that these LL energies are not sensitive to Rashba
SOC\cite{khooTunableQuantumHall2018}. Previous
experiments\cite{islandSpinOrbitdrivenBand2019,wangQuantumHallEffect2019} have
shown that one can quantify the Ising SOC $H_I = \frac{1}{2}\lambda_I\tau_z
s_z$ ($\lambda_I$ is the Ising SOC strength) with LLs on opposite graphene
layers: The sets of two Landau levels that cross at $\nu = \pm 3$ filling
factors have opposite layer polarization, such that their energy difference (at
zero $D$ field) is given by $\Delta E = E_Z \pm \lambda_I/2$ ($E_Z$ is the
Zeeman gap between spin-up and spin-down LLs)---only one of the two Landau
levels (with layer polarization close to the WSe$_2$) is affected by the Ising
SOC. Therefore, the critical field $B_\perp^*$ that makes $\Delta E$ vanish is
$2 E_Z = 2g\mu_B B_\perp^\ast =\lambda_I$. In \prettyref{exfig:QHE_ising}a-e$,
B_\perp^\ast \approx 3$~T is the magnetic field at which yellow and green
arrows level at the same $D$ field, yielding $\lambda_I \approx 0.7$~meV.

Independently, $\lambda_I$ can also be extracted from the doping-dependent FFT
splitting of quantum oscillations. \prettyref{exfig:QHE_ising}h inset shows the
FFT splitting $B_\text{split}$ as a function of doping at $D/\epsilon_0 =
0.2$~V/nm. Ising-type splitting is suppressed with increasing $|n|$, in
contrast to Rashba-type splitting which increases with increasing $|n|$. The
observed splitting is consistent with the value of $\lambda_I \approx 0.7$~meV
extracted from the quantum Hall measurements, as shown in the
\prettyref{exfig:QHE_ising}h inset by comparing to the band splitting predicted
from the band structure calculations at the same $D$ field. This method is,
however, less clean than the Landau level extraction, because Rashba SOC
additionally contributes to a spin splitting for both signs of $D$ (see below).
 
\textbf{Rashba SOC:} The effect of Rashba SOC is more subtle in the experiment.
Quantum oscillations at higher $B_\perp$ field provide an upper bound for the
magnitude of Rashba SOC. \prettyref{exfig:QHE_ising}f-i shows $\Delta R_{xx}$
versus $1/B_{\perp}$ and corresponding FFT measured at $D/\epsilon_0= 0.2$~V/nm
and $-0.1$~V/nm, respectively. At $D > 0$ (\prettyref{exfig:QHE_ising}h), FFT
reveals a frequency splitting while at $D < 0$ the splitting is absent
(\prettyref{exfig:QHE_ising}i). These observations are consistent with the
interpretation that at $D > 0$, the splitting is mainly caused by Ising SOC;
however at $D < 0$, the Ising effect is strongly diminished and Rashba SOC strength
$\lambda_R$ is not big enough to induce an observable splitting. The FFT peak
at $D < 0$ (\prettyref{exfig:QHE_ising}i) has a full width at half maximum
around $0.8$~T, which translates to an upper bound for the bare Rashba SOC
strength $\lambda_R \lesssim 5$ meV by comparing to the spin splitting
predicted from band structure calculations at the same density $n = -2\times
10^{12}~\text{cm}^{-2}$ and displacement field $D/\epsilon_0 = -0.1$~V/nm. An
upper bound on Rashba SOC can also be extracted from the observed spin
splitting at positive $D/\epsilon_0 = 0.2$~V/nm, assuming Ising SOC $\lambda_I =
0.7$~meV (see \prettyref{exfig:QHE_ising}h, inset). From this analysis we find
an upper bound $\lambda_R \lesssim 4$ meV, roughly consistent with the bound
from the negative $D$ field data.

\noindent {\bf Acknowledgments:} We thank
Andrea Young and Allan Macdonald for fruitful discussions. 
{\bf Funding:} This work has 
been primarily supported by NSF-CAREER award (DMR-1753306), and 
Office of Naval Research (grant no. N142112635), and Army Research 
Office under Grant Award W911NF17-1-0323. Nanofabrication efforts 
have been in part supported by Department of Energy DOE-QIS program 
(DE-SC0019166). S.N-P. acknowledges support from 
the Sloan Foundation (grant no. FG-2020-13716). J.A.
and S.N.-P. also acknowledge support of the Institute for 
Quantum Information and Matter, an NSF Physics Frontiers 
Center with support of the Gordon and Betty Moore Foundation 
through Grant GBMF1250. C.L. and E.L.H. acknowledge support from the 
Gordon and Betty Moore Foundation’s EPiQS Initiative, 
grant GBMF8682.

\noindent {\bf Author Contribution:} Y.Z. and S.N.-P. designed the experiment.
Y.Z., R.P. and H.Z. performed the measurements, fabricated the devices, and
analyzed the data. A.T., E.L.-H. and C.L. developed theoretical models and
performed calculations supervised by J.A. K.W. and T.T. provided hBN crystals.
S.N.-P. supervised the project. Y.Z., A.T., E.L.-H., C.L., H.Z., R.P., J.A.,
and S.N.-P. wrote the manuscript with the input of other authors.
 
\noindent{\bf Competing interests:} The authors declare no competing interests.

\noindent {\bf Data availability:} The data supporting the findings of this
study are available from the corresponding authors on reasonable request.
 
\noindent {\bf Code availability:} All code used in modeling in this study is
available from the corresponding authors on reasonable request.
 
\clearpage
\clearpage
\beginsupplement

\clearpage

\clearpage

\clearpage

\clearpage

\clearpage

\clearpage

\clearpage

\clearpage

\clearpage

\clearpage

{\large {\bf Supplementary Information:}} \\ 
{\large Spin-Orbit Enhanced Superconductivity in Bernal Bilayer Graphene} \\
Yiran Zhang, Robert Polski, 
Alex Thomson, Étienne Lantagne-Hurtubise, Cyprian Lewandowski, Haoxin Zhou, Kenji Watanabe, Takashi Taniguchi, 
Jason Alicea, and Stevan Nadj-Perge

{\bf Theoretical Analysis}

\section{Continuum model band structure of bilayer graphene}
\label{theory: continuum_model_BLG}

We consider the low-energy continuum model commonly used to describe
Bernal-stacked bilayer graphene
(BLG)\cite{mccannElectronicPropertiesBilayer2013}, under a perpendicular
displacement field $D$ which generates a potential difference $u = -d_\perp
D/\epsilon_\mathrm{BLG}$ between the top and bottom layers. Here
$d_\perp=0.33$~nm is the interlayer distance and $\epsilon_\mathrm{BLG} \sim
4.3$ is the relative permittivity of bilayer graphene. A continuum
approximation of the band structure returns a Hamiltonian of the form
 
\begin{align}
    \label{eqn:non-interacting_ham}
    H_0
    &=
    \sum_{\xi=\pm}\sum_\vk \psi^\dag_\xi(\vk)h_{0,\xi}(\vk)\psi_\xi(\vk),
&
    h_{0,\xi}({\vk}) &= \begin{pmatrix}
    u/2 &  v_0 \Pi^\dagger & -v_4 \Pi^\dagger & -v_3 \Pi \\
    v_0 \Pi & \Delta' +u/2 & \gamma_1 & -v_4 \Pi^\dagger \\
    -v_4 \Pi & \gamma_1 & \Delta' -u/2 & v_0 \Pi^\dagger \\
    -v_3 \Pi^\dagger & -v_4 \Pi & v_0 \Pi & -u/2
    \end{pmatrix}
\end{align}
where $\Pi = (\xi k_x + i k_y)$ and $v_i \equiv \frac{\sqrt{3} a}{2} \gamma_i$.
Here, $\xi=\pm1$ indicates the valley that has been expanded about: 
$\vK, \vK' = (\xi 4\pi/3a, 0)$ with $a = 0.246$~nm the lattice constant of 
monolayer graphene.
The  $4\times4$ matrix $h_\xi(\vk)$ is expressed in the sublattice/layer basis 
corresponding to creation/annihilation operators of the form 
$\psi_\xi(\vk)=\left( \psi_{\xi,A1}(\vk),\psi_{\xi,B1}(\vk),\psi_{\xi,A2}(\vk),\psi_{\xi,B2}(\vk)\right)^T$, 
where $A$/$B$ indicate the sublattice, $1$, $2$ indicate the layer, and the 
momentum $\vk$ is measured relative to $\vK_\xi$ (indices denoting the spin 
degrees of freedom have been suppressed).
It will sometimes be convenient below to express the Hamiltonian in terms of 
the spinors $\psi(\vk)=\left(\psi_{+}(\vk),\psi_-(\vk)\right)^T$.

The common values quoted for the five parameters entering the continuum model
in Eq.~\eqref{eqn:non-interacting_ham} are $\gamma_0 = 2.61$~eV (intralayer
nearest-neighbor tunneling), $\gamma_1 = 361$ meV (leading interlayer
tunneling), $\gamma_3 = 283$ meV (also known as trigonal warping term),
$\gamma_4 = 138$ meV, and $\Delta' = 15$ meV (potential difference between
dimer and non-dimer sites)\cite{jungAccurateTightbindingModels2014}.
 
A TMD monolayer adjacent to the graphene, such as is the case here with
WSe$_2$, is known to induce SOC via virtual
tunnelling\cite{gmitraTrivialInvertedDirac2016,gmitraProximityEffectsBilayer2017,
islandSpinOrbitdrivenBand2019}:
 
\begin{align}
    H_\mathrm{SOC}
    &=
    \sum_{\xi=\pm}\sum_\vk \psi^\dag_\xi(\vk) h_{\mathrm{SOC},\xi}\psi_\xi(\vk),
    &
    h_{\mathrm{SOC},\xi}(\vk)
    &=
    \mathcal{P}_1\left[
    \frac{\lambda_I}{2} \xi s^z 
    +
    \frac{\lambda_R}{2}\big(\xi\sigma^xs^y-\sigma^ys^x)
    \right],
\end{align}
where the Pauli matrices $\sigma^i$ and $s^i$, $i=x,y,z$, respectively act on
sublattice and spin degrees of freedom. The operator $\mathcal{P}_1$ projects
onto the top graphene sheet, i.e., only the sites A1 and B1:
$\mathcal{P}_1=\mathrm{diag}(\mathbb{1}_{2\times2},\mathbb{0}_{2\times2})$ in
the layer/sublattice basis used to express $h_{0,\xi}(\vk)$ in
\eqref{eqn:non-interacting_ham}. The parameters $\lambda_{I}$ and $\lambda_R$
quantify the strength of the Ising (also called ``valley-Zeeman'') and Rashba
SOC. Ab initio-type numerics and experimental estimates find a range of values
$\lambda_I\sim0-5$~meV and $\lambda_R\sim0-15$~meV for the SOC
parameters\cite{gmitraTrivialInvertedDirac2016,gmitraProximityEffectsBilayer2017,
wangOriginMagnitudeDesigner2016,yangStrongElectronholeSymmetric2017,
liTwistangleDependenceProximity2019, amannCounterintuitiveGateDependence2022},
which are also predicted to be strongly twist-angle
dependent\cite{liTwistangleDependenceProximity2019}.

In the absence of SOC and an applied displacement field with $v_3=v_4=0$, two
bands touch quadratically at charge neutrality. Two remaining bands are at
significantly higher and lower energies; their wavefunction are dominated by
the ``dimer sites,'' i.e., the A2 and B1 which sit immediately on top of one
another in the bilayer and hybridize strongly through the onsite tunnelling
parameter $\gamma_1$. Trigonal warping introduced by the $v_3$, $v_4$
associated hoppings in Eq.~\eqref{eqn:non-interacting_ham} splits the quadratic
band touching at charge neutrality into four distinct Dirac cones separated by
van Hove singularities (VHS). Turning on a displacement field $D$, a gap opens
at charge neutrality and the VHSs move apart in energy. Further, by flattening
the band bottom, the applied $D$ field also amplifies divergence of the DOS
close to the VHS. The low-energy states near $\vK$ and $\vK'$ become strongly
layer- and sublattice-polarized; e.g., on $A1$ sites for the valence band and
$B2$ sites for the conduction band, or vice versa for the other sign of $D$.
That is, the low-energy wavefunctions near charge neutrality and under a large
$D$ field are strongly localized on the ``non-dimer sites" of BLG.
 
The layer- and sublattice polarization of the low-energy wavefunctions near the
$\vK$, $\vK'$ points has important consequences for SOC induced by the TMD.
Indeed, Rashba SOC does not act effectively in the low-energy theory because it
is off-diagonal in the sublattice degree of freedom. It therefore induces a
splitting only at second order in degenerate perturbation theory, $\sim
\lambda_R^2/u$ with $u$ the interlayer potential (neglecting effects of further
perturbations such as trigonal warping, which we discuss below). By contrast,
the Ising SOC acts effectively in the subspace of sublattice- and
layer-polarized wavefunctions.
 
The normalized frequencies one expects from quantum oscillations for the
non-interacting theory are shown in \prettyref{exfig:theory_figure_cascade}a as
a function of hole doping for $D/\epsilon_0 = 1~\text{V/nm}$. The red lines
correspond to the spin-orbit-free case, whereas the blue lines are computed in
the presence of SOC; we additionally plot the Fermi surfaces for the spin-orbit
coupled band structure in the insets at a few representative fillings. The
Ising coupling is set to the experimentally extracted value
$\lambda_I=0.7$~meV, whereas for the Rashba coupling we select
$\lambda_R=3$~meV, which is within the upper bound consistent with the
experimental resolution. For the filling range shown, five different sets of
Fermi surface topologies are present for the case with SOC. For
$|n|<0.5\times10^{11}~\text{cm}^{-2}$, an $\mathrm{FP}(6)_{+}$ state is
realized in which each valley contributed three equally sized (but offset in
momentum space) Fermi surfaces. Assuming $\lambda_I>0$, given that the in-plane
mixing by Rashba is relatively small, the states that make up these pockets
largely have valley and spin quantum numbers $K',\uparrow$ and $K,\downarrow$.
With further doping, another six equally sized Fermi surfaces appear
corresponding to the spin degrees of freedom pushed down in energy by SOC
($K,\uparrow$ and $K',\downarrow$). This ${\rm FP}(6,6)_+$ state persists with
doping until $|n|\sim7.4\times10^{11}~\text{cm}^{-2}$, at which point the
system reaches a van Hove singularity and thus the Fermi surface structure
changes. The three majority pockets merge, resulting in two degenerate hole
pockets with a small electron pocket at their centre. With further doping the
electron-like pocket of the majority Fermi surfaces vanishes
($|n|\sim9\times10^{11}~\text{cm}^{-2}$). Subsequently, the minority pockets
reach the same VHS ($|n|\sim9.8\times10^{11}~\text{cm}^{-2}$) leading to the
formation of the small electron pocket.
 
A comparison of \prettyref{exfig:theory_figure_cascade}a with the quantum
oscillations data in \prettyref{fig:Fig3}c makes it clear that the
non-interacting theory is insufficient. In particular, SOC does explicitly
``polarize'' the band---they are energetically split---so that the ${\rm
FP}(6,6)_+$ phase is realized; this state persists up to
$|n|\sim7.4\times10^{11}~\text{cm}^{-2}$ at $D=1$~V/nm. By contrast, in the
experiment the ${\rm FP}(6,6)_+$ phase terminates at around
$|n|\sim5\times10^{11}~\text{cm}^{-2}$, where it is replaced by the ${\rm
FP}(2,2)_+$ state. Hence, while the Ising and Rashba parameters (obtained from
measurements at zero $D$ and doping) do arguably polarize the bands in the
non-interacting limit, the resulting splitting is not large enough to account
for the observed phase diagram.
 
\section{Interactions}
\label{theory: interactions}

The resistance data as a function of displacement field and doping clearly
demonstrate that the non-interacting band structure implied by
$H_0+H_\mathrm{SOC}$ in the previous section cannot fully describe the system.
Instead, given the large density of states close to charge neutrality in the
presence of large displacement fields, a series of polarized phases are
observed, which are naturally explained as a consequence of the Coulomb
interaction.
 
The Coulomb interaction is given by 
\begin{align}\label{eqn:Coulomb_Ham}
    H_C
    &=
    \frac{1}{2A}\sum_{\vk,\vk',\vq} V(\vq)\psi_\alpha^\dag(\vk)\psi_\beta^\dag(\vk')\psi_\beta(\vk'-\vq)\psi_\alpha(\vk+\vq).
\end{align}
Here the indices $\alpha$ and $\beta$ sum over valley, layer, spin, and sublattice degrees of freedom and $A=A_\mathrm{u.c.}N_\mathrm{site}$ is the total area of the sample, with $A_\mathrm{u.c.}$ denoting the unit cell area, $A_\mathrm{u.c.}=\sqrt{3}a^2/2$, and $N_\mathrm{site}$ denoting the total number of sites.
The unscreened Coulomb potential is $V(\vq)=e^2/(2\epsilon_r\epsilon_0 |{\vq}|)$, where $\epsilon_r\sim4.3$ is the dielectric constant for hBN-screened graphene.
Instead of considering this model, we look at a far simpler model in which the interaction is fully local: $V(\vq)=A_{\mathrm{u.c.}}U_C$.
We can roughly estimate 
\begin{align}
    U_C&\sim
    \frac{1}{A_\mathrm{u.c.}}\frac{e^2}{4\pi\epsilon_r\epsilon_0}d,
\end{align}
where we have substituted $2\pi/q_F\sim d$ with $d$ the inter-particle spacing.
A density of $n=-5\times10^{11}$~cm$^{-2}$ roughly translates to an inter-particle spacing of $d=15$~nm, which in turn implies $U_C\sim100$~eV. 
This estimate should be taken as an upper bound since it does not include the effects of screening.
Accordingly, more reasonable results are obtained by allowed the effective Coulomb interaction strength to take smaller values. 
In particular, we often select $U_C=35$~eV in accord with earlier calculations of Bernal stacked systems\cite{zondinerCascadePhaseTransitions2020,zhouIsospinMagnetismSpinpolarized2022}.

We emphasize that $U_C$ should \emph{not} be thought of as the setting the ``energy scale'' of the problem. 
Instead, the interactions naturally scale with the density.
In particular, if we let $\nu_f$ denote the number of electrons per unit cell, $\nu_f=A_\mathrm{u.c.}\cdot n=\sqrt{3}a^2/(2d^2)$, then the energy per electron is
\begin{align}\label{eqn:Coulomb_scale}
    \epsilon_C=\frac{E_C}{\mathrm{electron}}\sim \frac{e^2}{4\pi\epsilon_r \epsilon_0}\frac{1}{d},
\end{align}
which is precisely what we would have found with a real space description.
In this case, we find $\epsilon_C\sim20$~meV for $d=15$~nm.

Even with the long-range Coulomb form, $V(\vq)\propto1/|\vq|$, the interaction presented in Eq.~\eqref{eqn:Coulomb_Ham} is not fully general.
Instead, it was derived by taking the zero momentum portion of the density. 
In effect, the density can be expanded in terms of the continuum model operators as $\rho(\vr)\sim \rho_{++}(\vr)+\rho_{--}(\vr)+e^{-i\vK\cdot\vr}\rho_{-+}(\vr)+e^{i\vK\cdot\vr}\rho_{+-}(\vr)$ where $\rho_{\xi\xi'}(\vr) = \psi^\dag_\xi(\vr)\psi_{\xi'}(\vr)$ with $\psi_\xi(\vr)$ the real space version of the annihilation operator defined in SI, section~\ref{theory: continuum_model_BLG}.
Equation~\eqref{eqn:Coulomb_Ham} only includes $\rho_{++}$ and $\rho_{--}$, which accounts for the long-range part of the Coulomb interaction.
The Hund's term includes the remaining two piece of the density carrying momentum $\vK$, $\vK'$ and thus necessarily has a minimum momentum transfer of $\vK$.
Its magnitude can therefore be characterized by  $V(\vK)=e^2/(4\pi\epsilon_r\epsilon_0|\vK|)$.
Translating this scale into the relevant energy scale like in \eqref{eqn:Coulomb_scale}, we find
\begin{align}
    \epsilon_H\sim \frac{e^2}{4\pi\epsilon_r\epsilon_0}\frac{1}{a}\sim \frac{d}{a}\epsilon_C,
\end{align}
where factors of order unity have been neglected.

\section{Symmetries}
\label{theory: Symmetries}

We begin by discussing the flavour symmetries of the Hamiltonian in the absence of SOC.
It immediately follows that the system is invariant under the usual $\mathrm{SU}(2)_s$ spin rotation symmetry: $\psi(\vk)\to e^{i\theta \boldsymbol{n}\cdot \boldsymbol{s}/2}\psi(\vk)$, where $\boldsymbol{n}$ is an arbitrary unit 3-vector. 
The system similarly preserves the familiar $\mathrm{U}(1)_c$ phase rotation symmetry associated with charge conservation, $\psi(\vk)\to e^{i\theta}\psi(\vk)$.
These two standard symmetries are further augmented in bilayer graphene by the preservation of particle number individually within each valley, which follows from the so-called $\mathrm{U}(1)_v$ valley symmetry; its action takes the form $\psi(\vk)\to e^{i\theta\tau^z}\psi(\vk)$, where $\tau^z$ is a Pauli matrix acting on the valley indices of $\psi(\vk)$. 
In essence, the $\mathrm{U}(1)_v$ valley symmetry is a manifestation of the low energy scales at work: extrinsic scattering between states originating from valley $\vK$ to those originating from valley $\vK'$ are necessarily short range and thus precluded by the high quality of the sample.

Further inspection of the Hamiltonian $H_0+H_C$ reveals that the physical symmetry group, $\mathrm{U}(1)_c\times\mathrm{U}(1)_v\times\mathrm{SU}(2)_s$, is in fact a subgroup of a much larger \emph{effective} symmetry operative at the dominant energy scales of the system.
In particular, the Hamiltonian is invariant under independent \emph{spin} rotations within each valley: 
$\psi(\vk)\to \left( \mathcal{P}_+e^{i\theta_+\boldsymbol{n}_+\cdot\boldsymbol{s}/2}+\mathcal{P}_-e^{i\theta_-\boldsymbol{n}_-\cdot\boldsymbol{s}/2}\right)\psi(\vk)$, where $\boldsymbol{n}_\pm$ are unit 3-vectors and $\mathcal{P}_\pm$ project onto the valley $K$ and $K'$.
Together with the two $\mathrm{U}(1)$ symmetry groups, the result is the existence of a $\mathrm{U}(2)_K\times\mathrm{U}(2)_{K'}\cong\mathrm{U}(1)_c\times\mathrm{U}(1)_v\times\mathrm{SU}(2)_K\times\mathrm{SU}(2)_{K'}$ flavour symmetry.
Importantly, this result implies that any degeneracies encoded by the $\mathrm{U}(2)_K\times\mathrm{U}(2)_{K'}$ effective symmetry are split only at the scale of the Hund's coupling, $\epsilon_H\sim (a/d)\epsilon_c$.

The introduction of SOC naturally reduces both the effective and physical symmetry groups. 
The Ising term by itself ($\lambda_R=0$) reduces the $\mathrm{SU}(2)_s$ to $\mathrm{U}(1)_z$, the group generating rotations about the spin-$z$ axis, and the remaining physical symmetry group is thus $\mathrm{U}(1)_c\times\mathrm{U}(1)_v\times\mathrm{U}(1)_z$.
The large effective symmetry group relevant to scales larger than $\epsilon_H$ is similarly diminished by the restriction that only spin-$z$ rotations in either valley leave Hamiltonian unmodified.
The result is an effective symmetry group composed of four different $\mathrm{U}(1)$ rotations: $\mathrm{U}(1)_c\times\mathrm{U}(1)_v\times\mathrm{U}(1)_{z,K}\times\mathrm{U}(1)_{z,K}$, where $\mathrm{U}(1)_{z,K^{(\prime)}}$ rotates the spin of the valley $K^{(\prime)}$ fermions  about the $z$-axis.

When Rashba spin-orbit coupling is present, with or without Ising SOC, all global, continuous spin rotations are absent. 
Both the physical and effective flavour symmetry groups are pared down to $\mathrm{U}(1)_c\times\mathrm{U}(1)_v$. 
The small upper bound for the Rashba coupling $\lambda_R$ imposed by experiment leads us to largely neglect its symmetry breaking effect. 

The Hamiltonian also possesses a number of discrete symmetries, the most important of which is time reversal symmetry (TRS):
\begin{align}
    \mathcal{T}&:
    &
    \psi(\vk)&\to i\tau^xs^y\psi(-\vk),
    &
    i&\to-i.
\end{align}
Time reversal remains a good symmetry of the system both with and without spin-orbit coupling.

\section{Mean field approximation}
\label{theory: Mean_field_approximation}

We study the interacting theory using mean field theory.
In particular, for each filling $\nu$, where $\nu$ is the number of carriers per unit cell as measured relative to charge neutrality, we find the Slater determinant ground state $|\phi_\nu\rangle$ that minimize the mean-field ground state energy $E^{(\nu)}_\mathrm{MF}[\phi]=\langle \phi_\nu |\left(H_0+H_\mathrm{SOC}+H_C\right)|\phi_\nu\rangle$. 
This procedure is essentially equivalent to replacing the interacting Hamiltonian with the one-particle mean field Hamiltonian
\begin{align}\label{eqn:MF-hamiltonian}
    H^{(\nu)}_\mathrm{MF}
    &=
    \sum_\vk \psi^\dag(\vk)h_\mathrm{MF}\psi(\vk),
    &
    h^{(\nu)}_\mathrm{MF}
    &=
    -\frac{U_C}{N_\mathrm{site}}
    \sum_{\vq}
    \Big[
    P^{(\nu)}(\vq)
    -
    \mathrm{tr}\big(P^{(\nu)}(\vq)\big)\,\mathbb{1}
    \Big],
\end{align}
where the projector $P(\vk)$ is given by
\begin{align}\label{eqn:P-def}
    P^{(\nu)}_{\alpha\beta}(\vq)
    &=
    \langle \psi^\dag_\beta(\vq)\psi_\alpha(\vq)\rangle_\nu
    =
    \langle \psi^\dag_\beta(\vq)\psi_\alpha(\vq)\rangle
    -
    \langle \psi^\dag_\beta(\vq)\psi_\alpha(\vq)\rangle_\mathrm{CNP}.
\end{align}
Here, the values of the correlation functions $\langle\cdot\rangle$ are in turn obtained by diagonalizing $H_0+H_\mathrm{SOC}+H_\mathrm{MF}^{(\nu)}$, and the subscript ``CNP'' indicates that the expectation value is being taken with respect to the charge neutrality point.
Self-consistency is attained when the mean field term $H_\mathrm{MF}^{(\nu)}$ used to calculate $P^{(\nu)}$ is in turn defined via Eq.~\eqref{eqn:MF-hamiltonian}.
It can be shown that the ground state of this self-consistent Hamiltonian is a local minimum of the mean field energy functional $E_\mathrm{MF}^{(\nu)}$.

We solve for $P^{(\nu)}$ through the following procedure. 
We select an initial value $H_{\mathrm{MF}}^{\mathrm{init}.}$ and then iterate between Eqs.~\eqref{eqn:MF-hamiltonian} and~\eqref{eqn:P-def} until self-consistency is reached.
Crucially, states possessing less symmetry than the initial Hamiltonian $H_0+H_\mathrm{SOC}+H_{\mathrm{MF}}^{\mathrm{init}.}$ are inaccessible.
For instance, if the initial mean field Hamiltonian is invariant under the $\mathrm{U}(1)_v$ symmetry, the final wavefunction $|\phi_\nu\rangle$ (and the corresponding $P^{(\nu)}$) must also be invariant under the $\mathrm{U}(1)_v$ symmetry and hence so must $H_\mathrm{MF}^{(\nu)}$.
As noted, the symmetries allow us to separate the problem into those that preserve the $\mathrm{U}(1)_v$ and those that break it.

In principle, the result should be the minimal energy state that respects the same symmetries as $H_\mathrm{MF}^{\mathrm{init.}}$ and the non-interacting terms, $H_0+H_\mathrm{SOC}$.
However, in practice, the algorithm sketched above sometimes finds itself trapped in local minima,  unable to attain the true ground state within that symmetry class.
This happenstance is particularly common when there are many nearly degenerate ground states, which, as we describe in the following section, is the case here.
We have not rigorously explored the phase diagram to ensure that all of the solutions presented below represent true symmetry-class ground states largely because the simplicity of the model makes it more appropriate for a qualitative study of trends, as opposed to a quantitative one.
There is therefore little reason to ignore low energy states in favour of what, according to this model, is the ``true'' ground state.
In fact, the phenomenological arguments we make below in SI, section~\ref{theory: Ising_SOC_perturbation} imply that a different ground state is realized than suggested by our simulations.

We are primarily interested in what happens upon hole doping the system in the presence of a positive displacement field, and we therefore specialize to this scenario; our discussion can readily be translated to the case with opposite $D$-field sign as well as with electron doping. 
We further note that provided the $D$-induced gap at charge neutrality is sufficiently large, we do not expect $H_C$ to induce significant mixing between the four sets of (effectively) degenerate spin-valley bands defined by $H_0+H_\mathrm{SOC}$, allowing us to restrict our focus entirely to the active bands of interest. 

While the mean field Hamiltonian $h_\mathrm{MF}^{(\nu)}$ is independent of momentum, it nevertheless acts on the original 16 degrees of freedom as opposed to the four bands of interest.
It is convenient to distill the resulting Hamiltonian to the only degrees of freedom that remain upon projecting to the bands of interest. 
In particular, instead of directly discussing $h_\mathrm{MF}^{(\nu)}$, we focus instead on
\begin{align}\label{eqn:simple_mf_ham}
    h^{(\nu)\prime}_\mathrm{MF}
    &=
    \sum_{\substack{a,i=0,x,y,z\\(a,i)\neq(0,0)}}t_{ai}\tau^as^i,
    &
    t_{ai}&=\frac{1}{4}\mathrm{tr}\big( h_\mathrm{MF}^{(\nu)}\tau^as^i \big),
\end{align}
where we do not include the constant shift of the chemical potential.
Notably, $h^{(\nu)\prime}_\mathrm{MF}$ is still a $16\times16$ matrix, but with any dependence on either the layer or sublattice removed.
It follows that $h^{(\nu)\prime}_\mathrm{MF}$ does not account for some of the spatial dependence that results when one projects onto the four hole bands close to charge neutrality.
These effects, while present in our numerics, are largely irrelevant for the purpose of understanding the resulting phases.

\section{Polarized phases}
\label{theory: Polarized_phases}

We are most interested here in the spontaneous breaking of the spin-valley symmetries, resulting in the polarized states seen in experiment.
The propensity for this type of symmetry breaking follows from the large density of states induced by the displacement field.
Interactions make having a large density of states at the Fermi energy energetically costly. 
At the expense of the kinetic energy, the DOS at the Fermi energy and its associated energy cost may be lowered by breaking the flavour symmetry and alternately increasing and decreasing the filling of certain flavours.
The advantage of this process is roughly encapsulated in the Stoner criterion, which states that polarization occurs when $V\rho\geq1$, where $V$ is the interaction scale and $\rho$ the density of states.

At the mean field level, the polarized phases are characterized by the (simplified) mean field Hamiltonian ${h}^{(\nu)\prime}_\mathrm{MF}$ of Eq.~\eqref{eqn:simple_mf_ham}.
We begin by addressing the phases in the absence of SOC where the effective $\mathrm{U}(2)_K\times \mathrm{U}(2)_{K'}$ symmetry remains a good approximation.
In the simplest case, only a single $t_{ai}$ in Eq.~\eqref{eqn:simple_mf_ham} is non-zero:
\begin{align}\label{eqn:singly_pol}
    {h}^{(\nu)\prime}_\mathrm{MF} = t_{ai} \tau^as^i.
\end{align}
This mean field term pushes two flavours up and two flavours down in energy, resulting in a set of minority and a set of majority Fermi pockets. 
In what follows, this type of phase is denoted ``singly polarized.''
Such singly polarized phases are not limited by mean field Hamiltonians of the form Eq.~\eqref{eqn:singly_pol}, but are more generally induced by any $h^{(\nu)\prime}_\mathrm{MF}$ given by a sum of \emph{anticommuting} matrices $\tau^as^i$.

We group the singly polarized phases resulting from Eq.~\eqref{eqn:singly_pol} into two broad categories.
First are the ``simple'' polarized states that preserve the $\mathrm{U}(1)_v$ valley symmetry, implying that the mean field Hamiltonian associated with such states satisfies $[\tau^z,{h}^{(\nu)\prime}_\mathrm{MF}]=0$:
\begin{align}
    \tilde{h}^{(\nu)}_\mathrm{MF}&\propto \tau^as^i
    &
    &\text{where}
    &
    \tau^as^i&\in\{ s^{x,y,z},\,\tau^zs^{0,x,y,z}\}.
\end{align}
Notably, the ${h}^{(\nu)\prime}_\mathrm{MF}$ above commutes with the non-interacting  Hamiltonian $h_0(\vk)$, and it follows that its primary effect is to generate a relative shift of the band energies.
The action of $\mathrm{U}(2)_K\times\mathrm{U}(2)_{K'}$ rotates the orders leading to spin polarization (SP) (${h}^{(\nu)\prime}_\mathrm{MF}\propto s^{x,y,z}$) and spin-valley polarization (SVP) (${h}^{(\nu)\prime}_\mathrm{MF}\propto \tau^zs^{x,y,z}$) into each other, hence these states must have the same energy (with respect to the Hamiltonian under consideration currently).
Similarly, the action of $\mathrm{U}(2)_K\times\mathrm{U}(2)_{K'}$ may also rotate ${h}^{(\nu)\prime}_\mathrm{MF}$ to a linear combination of these order parameters, for instance to induce spin polarization along an arbitrary direction.
By contrast, the valley polarized (VP) state characterized by ${h}^{(\nu)\prime}_\mathrm{MF}\propto \tau^z$ does not break $\mathrm{U}(2)_K\times\mathrm{U}(2)_{K'}$ (although it does break time reversal), and it is therefore not necessarily degenerate with the SP and SVP states.
However, the density-density form of the Coulomb interaction renders this distinction meaningless and prevents the system from distinguishing whether valley $K$ or $K'$ is filled on average. 
Additional interaction terms---say arising from short-range interactions---will split this accidental degeneracy.
For instance, the phonon interaction\cite{chouAcousticphononmediatedSuperconductivityBernal2022} takes the form $\sim\int_\vr (\psi^\dag\tau^z\psi)^2$ and thus both preserves the $\mathrm{U}(2)_K\times\mathrm{U}(2)_{K'}$ symmetry while clearly distinguishing between VP and SP/SVP states.
(The Hund's term whose inclusion does decrease the effective symmetry group also distinguishes these two sets of states.)

The second category of states breaks the $\mathrm{U}(2)_K\times\mathrm{U}(2)_{K'}$ to a \emph{diagonal} subgroup through the spontaneous generation of inter-valley tunnelling.
These ``inter-valley coherent'' (IVC) ordered states occur when $[\tau^z,{h}^{(\nu)\prime}_\mathrm{MF}]\neq0$:
\begin{align}
    \tilde{h}^{(\nu)}_\mathrm{MF}&\propto \tau^as^i
    &
    &\text{where}
    &
    \tau^as^i\in\{ \tau^xs^{0,x,y,z}, \tau^ys^{0,x,y,z}\}.
\end{align}
As with the SP and SVP states, the IVC order parameters may all be mapped to one another through the action of $\mathrm{U}(2)_K\times\mathrm{U}(2)_{K'}$, meaning that they must be degenerate.
Unlike the previous set of states, the IVC mean field Hamiltonian mixes states from different valleys and therefore significantly alters the form of the band structure.

In addition the singly polarized states, the system may also favour breaking more than a single symmetry, resulting in a ``multiply polarized'' state.
In this case, ${h}^{(\nu)\prime}_\mathrm{MF}$ is a sum of multiple \emph{commuting} $\tau^as^i$ matrices.
An example of such a mean field term is
\begin{align}
    {h}^{(\nu)\prime}_\mathrm{MF}&= t_0\tau^z + t_1 s^z + t_3\tau^zs^z.
\end{align}
When $|t_1|=|t_2|=|t_3|$, this mean field Hamiltonian pushes one flavour to a higher or lower energy on average, leaving the remaining three degenerate.
More commonly, however, we find that the coefficients satisfy $|t_1|>|t_2|\cong|t_3|$.
As with the ``singly polarized'' state above, the multiply polarized states may also be categorized depending on whether they break or preserve $\mathrm{U}(1)_v$. 

We calculated the self-consistent mean field Hamiltonians and corresponding ground states in the absence of SOC for a variety of parameters.
For filling ranges that prefer singly polarized states, we consistently find IVC ordered states to have the lowest energies.
In this case, no other symmetries are broken.
Similarly, for filling ranges preferring multiply polarized states, IVC order is typically also generated, although this time it must be present alongside another symmetry breaking order.
We stress, however, that our model is very crude and is not expected to yield quantitatively accurate results.

We finally turn to the case of primary interest: bilayer graphene with proximity-induced SOC.
Given the relative smallness of the effective Rashba coupling in the parameter range of interest, we focus on a system with only Ising SOC; modifications brought by the reintroduction of Rashba are briefly addressed below. 
In this case, the $\mathrm{U}(2)_K\times\mathrm{U}(2)_{K'}$ is reduced to a $\mathrm{U}(1)_c\times\mathrm{U}(1)_v\times \mathrm{U}(1)_{K,z}\times\mathrm{U}(1)_{K,z'}$, where $\mathrm{U}(1)_c$ and $\mathrm{U}(1)_v$ denote the charge and valley symmetries while $\mathrm{U}(1)_{K^{(\prime)},z}$ represents spin rotations about the $z$ axis in either valley.
It naturally follows that the $\mathrm{SU}(2)_s$ spin degeneracies present in the absence of SOC are lifted. 
In fact, because the Ising SOC term resembles a self-generated ``spin-valley order'', $\tau^zs^z$, its presence results in a `singly polarized' state even in the non-interacting limit. 
(As discussed in SI, section~\ref{theory: continuum_model_BLG} and shown in \prettyref{exfig:theory_figure_cascade}a, the splitting induced by the bare Ising coupling is substantially smaller than what is required to explain the phases seen in experiment.) 
Clearly, \emph{if} the SP or SVP phases were the preferred ground state without SOC, the mean field Hamiltonian generated in the presence of SOC would have a clear energetic preference for Ising-like SVP polarized states.
Indeed, we consistently find that the effective Ising SOC is enhanced by the interactions.

The introduction of Rashba SOC breaks the $\mathrm{SU}(2)$ spin symmetries operative in either valley, leaving only a $\mathrm{U}(1)_c\times\mathrm{U}(1)_v$ flavour symmetry. 
Although it is included below, it has relatively little qualitative effect on the resulting phase diagrams.

In \prettyref{exfig:theory_figure_cascade}c,e we present the normalized frequencies expected in quantum oscillations and the corresponding Fermi surfaces obtained through simulations performed with $\lambda_I=0.7$~meV and $\lambda_R=3$~meV.
Before describing these results in detail, we emphasize that c and e were both simulated using a single choice of $H_\mathrm{MF}^\mathrm{init.}$.
As described in SI, section~\ref{theory: Mean_field_approximation}, although all phases represented in \prettyref{exfig:theory_figure_cascade}c,e are \emph{low energy} states, it is possible that our algorithm has not found the \emph{true} ground state of the model.
Since the simplicity of the model prevents us from making quantitative predictions based on its behaviour, we do not view this as a particularly significant failing of the simulations. 
Nevertheless, we have verified by doing multiple runs with different starting positions that the phases given in \prettyref{exfig:theory_figure_cascade}c,e are not flukes of our specific choice of $H_\mathrm{MF}^\mathrm{init.}$ but are instead overall representative of the different low energy states present.

The plot in \prettyref{exfig:theory_figure_cascade}c was obtained in the absence of $\mathrm{U}(1)_v$ breaking (as explained in SI, section~\ref{theory: Mean_field_approximation}, the presence of $\mathrm{U}(1)_v$ is enforced by our choice of $H_{\mathrm{MF}}^{\mathrm{init.}}$).
Comparing with \prettyref{exfig:theory_figure_cascade}a, it is clear even for low dopings, in the $\mathrm{FP}(6,6)_+$ phase, that the splitting between the two sets of pockets is larger when SOC is present: the simulation that included interactions finds minority Fermi surfaces that are even smaller relative to the majority pockets than what is seen without interactions.
Further doping sees a first order transition at around $|n|\sim5\times10^{11}\text{ cm}^{-2}$ where the ground state discontinously jumps to an $\mathrm{FP}(1,3,6)_+$ state, whose background is coloured yellow. 
Here, we see that the Fermi surfaces of the two majority flavours differ not only from the two minority flavours, but they also differ from one another.
The $\mathrm{FP}(1,3,6)_+$ state is therefore multiply polarized: in addition to the Ising polarization, an additional symmetry breaking order was generated (here, a mixture of spin $s^z$ and valley $\tau^z$ polarization).
When the doping reaches $|n|\sim5.8\times10^{11}\text{ cm}^{-2}$, another first order transition occurs, yielding a singly polarized state with two large and six small Fermi pockets, $\mathrm{FP}(2,6)_+$ (shown in red). 
Its development simply follows from a large enhancement of the effective $\lambda_I$; no additional symmetries are broken.
Another transition occurs at $|n|\sim8.8\times10^{11}\text{ cm}^{-2}$, into the $\mathrm{FP}(3,3)_+$ phase, where three of the flavours have large Fermi surfaces and one of the flavours has three small surfaces. 
Again, this state is multiply polarized, thus requiring additional symmetry breaking. %

Comparing the theory simulation of \prettyref{exfig:theory_figure_cascade} against the experimental data, it is tempting to associate the $\mathrm{FP}(2,6)_+$ phase found here with the experimentally observed $\mathrm{FP}(2,2)_+$ phase that serves as a parent to superconductivity, despite the difference in the number of small Fermi pockets.
The latter discrepancy may be justified through the subdominant inclusion of rotational symmetry breaking, which could spontaneously reduce the number of filled small pockets from six to two (we address this process in more detail in the next section as well as in \prettyref{exfig:theory_figure_cascade}d,f). 
However, as mentioned, the large enhancement of $\lambda_I$ required to obtain this phase is at odds with the observed in-plane magnetic field dependence\footnote{One may argue that if Rashba SOC were also enhanced by interactions, the Pauli-limit violation ratio could remain unchanged. However, we see no evidence in our calculations of any Rashba enhancement.}, making this type of Ising-dominated polarized phase an unlikely candidate.
The quantum oscillations characterizing the $\mathrm{FP}(3,3)_+$ phase is also reminiscent of the large-doping regime adjacent to the superconducting $\mathrm{FP}(2,2)_+$ phase. 
In particular, the downward sloping frequency around $\sim1/3$ is also present in \prettyref{fig:Fig3}c and \prettyref{exfig:Fan and FFT}
(Our assertion that the Ising polarized phase $\mathrm{FP}(2,6)_+$ is unlikely present experimentally does not rule out the experimental relevance of $\mathrm{FP}(3,3)_+$).

\prettyref{exfig:theory_figure_cascade}e shows the quantum oscillation frequencies and Fermi surfaces for a mean field solution defined with the same parameters as in c, but whose initialization condition allowed IVC order to develop.
Setting aside technicalities surrounding the self-consistent mean field procedure, we emphasize that IVC should technically only develop when it is energetically favourable to do so.
Unsurprisingly, the solution at low dopings is identical to what is shown in c, with only Ising SOC present.
A multiply polarized phase coloured in yellow, $\mathrm{FP}(1,6)_+$, is attained around $|n|\sim3.8\times10^{11}\text{ cm}^{-2}$, and the Fermi surface shape makes the difference between this solution and the one in c apparent.
While the two minority pockets shown in the inset resemble those found in the non-interacting and $\mathrm{U}(1)_v$-preserving cases (\prettyref{exfig:theory_figure_cascade}a,c), the large pocket is quite different---a direct consequence of the inter-valley hybridization.
A complicated series of intermediate phases existing only within a narrow filling range follows with additional doping before the system enters a singly polarized $\mathrm{FP}(2)_+$ phase at $|n|\sim6\times10^{11}\text{ cm}^{-2}$. 
Two large, star-shaped Fermi surfaces that clearly do not resemble those found in the interaction-free band structure are present, yet again as a direct consequence of IVC order; we colour this region in blue to to distinguish it from the red singly polarized phases without IVC order.
Despite its singly polarized nature, the IVC order responsible for the star-shaped Fermi surface is generated alongside an enhancement of the Ising order:
\begin{align}\label{eqn:IVC_SVP_pol}
    h_\mathrm{MF}^{(\nu)\prime} = \frac{\Delta_\mathrm{IVC}}{2} \tau^x + \frac{\delta \lambda_I}{2} \tau^zs^z.
\end{align}
Importantly, the state remains singly polarized because $\tau^x$ and $\tau^zs^z$ anticommute: only a single gap is opened by the mean field potential.
Further doping leads first to a $\mathrm{U}(1)_v$-breaking multiply polarized state, then to a $\mathrm{U}(1)_v$-preserving polarized state analogous to what is realized at the same filling range in \prettyref{exfig:theory_figure_cascade}c, and then finally to another IVC-ordered multiply polarized state.

Unlike the SOC-free model, where IVC order was always found to have the lowest energy, the addition of Ising SOC has made the SVP phase competitive against the IVC---even when IVC order was allowed to develop, there is a least one region in \prettyref{exfig:theory_figure_cascade}e where the Ising SVP state is preferred. 
In fact, the effective Ising coupling is substantially enhanced both in the singly polarized state without IVC order in \prettyref{exfig:theory_figure_cascade}c and with IVC order in \prettyref{exfig:theory_figure_cascade}e---more than is strictly compatible with the in-plane field measurements of the superconducting state. 
However, importantly, the latter state has spontaneously broken an additional symmetry relative to the Ising SOC-induced SVP order, 
establishing it as a distinct phase. 
Within this IVC-ordered state, we expect the relative magnitudes of the effective Ising SOC term and the IVC order $\Delta_\mathrm{IVC}$ is a matter of details---precisely the quantitative information our model in unable to provide reliably.

\section{Nematicity}
\label{theory: Nematicity}
In addition to the internal flavour symmetries of the continuum model, the system also possess a $C_3$ symmetry that rotates the system by 120\degree. 
This transformation acts on the spinors $\psi_\xi$ as 
\begin{align}
    C_3&:\qquad
    \psi_\xi(\vk)
    \to 
    e^{i 2\pi \xi\sigma^z/3}\psi_\xi(R_3\vk),
    &
    R_3
    &=
    \left(\begin{matrix}
    -1/2    &   \sqrt{3}/2  \\ -\sqrt{3}/2  &   -1/2
    \end{matrix}\right).
\end{align}
As alluded to in the discussion of \prettyref{exfig:theory_figure_cascade}c,
the experimental observation of the $\mathrm{FP}(2,2)_+$ state is consistent with a spontaneous breaking of this rotational symmetry, i.e., nematicity.
The development of nematic order had been predicted in this system through a momentum-condensation-like Pomeranchuk instabilities\cite{jungPersistentCurrentStates2015,dongIsospinFerromagnetismMomentum2021, huangSpinOrbitalMetallic2022}. %
Reference~\citenum{dongIsospinFerromagnetismMomentum2021} argues that the development of the nematic order is subdominant to the polarizing energy scale. 
That is, the internal flavour symmetries are first broken in the manner described in the previous section, and the electrons subsequently choose to occupy one out of the three small pockets instead of occupying all three pockets equally.
The momentum space dependence of the Coulomb interaction, which our simulations ignore, plays a crucial role in the derivation of this effect, and our model is therefore unable to self-consistently prefer the formation of the nematic order.

To compensate for the lack of spontaneous nematic order, we instead explicitly break the $C_3$ symmetry by modifying the non-interacting portion of the Hamiltonian.
In particular, we replace $h_{0,\xi}(\vk)$ in Eq.~\eqref{eqn:non-interacting_ham} with $h_{0,\xi}(\vk)+\delta h_\mathrm{nem}$ where
\begin{align}
    \delta h_\mathrm{nem}
    &=
    \begin{pmatrix}
    0   &   0   &   \alpha_1 + \alpha_2 &   \alpha_3  \\
    0   &   0   &   0   &   -\alpha_1 + \alpha_2  \\
    \alpha_1 + \alpha_2   &   0   &   0   &   0 \\
    \alpha_3 &   -\alpha_1 + \alpha_2   &   0   &   0  
    \end{pmatrix}.
\end{align}
We otherwise implement the identical procedure to the one described in the previous section, with the results shown in \prettyref{exfig:theory_figure_cascade}d and~f.

The simulations responsible for \prettyref{exfig:theory_figure_cascade}d are
the analogue to those of \prettyref{exfig:theory_figure_cascade}c in that the
$\mathrm{U}(1)_v$ symmetry was not allowed to break spontaneously.
Unsurprisingly, the explicit breaking of the $C_3$ symmetry makes the resulting
quantum oscillation frequencies and Fermi surface structures more complicated
than those of shown in \prettyref{exfig:theory_figure_cascade}c. Again, the low
doping regime is characterized by an enhancement of the effective Ising SOC
compared to the non-interacting theory, resulting in the $\mathrm{FP}(2,2,4)_+$
phase shown. Around $|n|\sim3.2\times10^{11}\text{ cm}^{-2}$, the system
transitions to a multiply polarized phase $\mathrm{FP}(1,1,2,2)_+$. This phase
undergoes a Lifshitz transition that does not change the polarizing order at
$|n|\sim5\times10^{11}\text{ cm}^{-2}$, after which it evolves continuously
into a singly ordered $\mathrm{FP}(2,2)_+$ phase at
$|n|\sim5.5\times10^{11}\text{ cm}^{-2}$. This $\mathrm{FP}(2,2)_+$ phase and
its higher doping partner $\mathrm{FP}(2,2,4)_+$ phase are the analogues of the
$\mathrm{FP}(2,6)_+$ phase in \prettyref{exfig:theory_figure_cascade}c in
accordance with the discussion of the previous section: these singly polarized
states do not arise out of an interaction-induced spontaneous breaking of a
symmetry, but simply out of the enhancement of the Ising SOC induced SVP order
(the enhancement, however, is so large relative to the scale of the Ising SOC
parameter appearing in the non-interacting Hamiltonian that it is still
reasonable to identify $\mathrm{FP}(2,2)_+$ and $\mathrm{FP}(2,2,4)_+$ as
polarized states). As in that section, however, we are forced to conclude that
the $\mathrm{FP}(2,2)_+$ phase here is not compatible with the in-plane field
measurements since such an extreme increase in the effective Ising SOC would
imply a far greater Pauli-limit violation than observed experimentally.
 
The plot in \prettyref{exfig:theory_figure_cascade}f illustrates a set of
solutions in which the ground state was allowed to develop IVC order. The low
doping regime is identical to d, but the multiply polarized phase the system
transitions into around $|n|\sim3.2\times10^{11}\text{ cm}^{-2}$ differs: the
star-like shape of the Fermi surface in the $\mathrm{FP}(1,2)_+$ phase clearly
indicates that IVC order is present. After some minor changes in the Fermi
surface topology, a first order transition to a singly polarized phase,
$\widetilde{\mathrm{FP}}(2,2)_+$, occurs at around $|n|\sim
4.6\times10^{11}\text{ cm}^{-2}$. This phase once more follows from the
enhancement of the effective Ising coupling as opposed to the spontaneous
breaking of an additional symmetry, and is thus partner to the
$\mathrm{FP}(2,2)_+$ phase in d. We note that the
$\widetilde{\mathrm{FP}}(2,2)_+$ in f appears at lower fillings than the same
phase appears in d. Since all phases realized in d can also be realized in f,
we would normally expect any phase lacking IVC in f to be represented in d at
those same filling. This discrepancy is related to the discussion of SI,
section~\ref{theory: Mean_field_approximation} on how the algorithm may find
local minima instead of true minima when many states with similar energies are
present. The existence of this near-degenerate manifold of mean field states is
apparent upon further doping, which sees the simulation alternate between
$\mathrm{U}(1)_v$ symmetric (coloured red) and $\mathrm{U}(1)_v$ breaking
(coloured blue) singly polarized ground states multiple times, up until
$|n|\sim8\times10^{11}\text{ cm}^{-2}$ where the system becomes multiply
polarized. Importantly, from within the IVC ordered $\mathrm{FP}(2)_+$ state, a
first order transition to a different IVC-ordered state possessing two large
and two small Fermi surfaces occurs at $|n|\sim5.5\times10^{11}\text{
cm}^{-2}$, reminiscent of the state that gives rise to superconductivity in the
experiment. 

With regards to the singly polarized states, we make no
claims that our model strictly prefers one of these options in the density
range shown relative to the other. The primary takeaway message from the this
discussion is that with nematicity, IVC-ordered $\mathrm{FP}(2,2)_+$
\emph{can be} realized in the system.

\section{Ising-SOC-mediated ground state selection}
\label{theory: Ising_SOC_perturbation}

These experiments prompt important questions regarding the role of WSe$_2$ and
the concomitant spin-orbit coupling in promoting superconductivity: what does
the SOC change so that bilayer graphene is able to superconduct at zero field?
The persistence of superconductivity across the entirety of the
$\mathrm{FP}(2,2)_+$ phase suggests the realization of this ground state as the
key to the development of superconductivity. We propose that in the absence of
SOC, interactions favour a ground state that is inhospitable to zero-field
superconductivity; with the addition of SOC, a distinct ground state amenable
to superconductivity is selected instead. Given the relatively small effect of
Rashba spin orbit on the band structure, we completely ignore its influence on
the interacting ground state selection for this discussion, focusing instead on
the effects of Ising SOC.
 
We first recall that Ising SOC itself splits the flavour degeneracy, resulting
in a non-interacting singly polarized SVP phase even at the level of the band
structure. We therefore infer that \emph{if} the SOC-free $\mathrm{FP}(2,2)_+$
ground state was an SVP polarized state, the addition of Ising SOC would not
alter the ground state. An SVP SOC-free ground state is therefore unlikely.
 
A natural next proposition is that the $\mathrm{FP}(2,2)_+$ phase seen experimentally in this paper \emph{is} precisely an Ising SVP state: the Coulomb interaction serves to increase the magnitude of SOC-induced band splitting, but otherwise induces no spontaneous symmetry breaking.
The numerically obtained phases shaded in red in \prettyref{exfig:theory_figure_cascade}c-d are all examples of such states.
Importantly, these states result from a large enhancement of the Ising coupling by interactions and are thus seemingly only consistent with a correspondingly enhanced Pauli-limit violation.
While the PVR is reasonably large in the low doping regime of the superconductor, at large dopings, the Pauli limit is barely violated at all.
This large variation in behaviour across the superconducting dome may be generally interpreted in two ways:
\begin{enumerate}[topsep=-2ex,itemsep=0ex,partopsep=-1ex,parsep=0pt]
    \item The nature of the interacting ground state $\mathrm{FP}(2,2)_+$ remains largely unchanged as a function of doping. 
    The change in PVR instead follows from relatively small
    band structure effects compared to the interaction scale, such as Rashba SOC or the orbital coupling of an in-plane magnetic field.  
    The specifics of this mechanism are discussed in more detail in the subsequent section.
    \item The nature of the interacting ground state changes substantially as a function of a doping.
\end{enumerate}
These two possibilities are of course not mutually exclusive nor even strictly distinct. 
They nevertheless provide a useful framework for organizing the energy scales and their implications in what follows.

Taking the perspective of scenario (1), we conclude that the $\mathrm{FP}(2,2)_+$ is incompatible with an Ising-induced SVP ground state.
We are thus left with a scenario in which
Ising SOC selects a non-SVP phase that in turn hosts superconductivity. 
Similarly, the same reasoning used to reject the SVP states removes the SP states as potential ground states---the Pauli-limit violation of an SP state would be even larger than expected for an SVP state. Assuming the full $\mathrm{U}(2)_K\times\mathrm{U}(2)_{K'}$ symmetry, the discussion in SI, section~\ref{theory: Polarized_phases} leaves two remaining classes of singly polarized ground states: the VP state
($h^{(\nu)\prime}_\mathrm{MF}\propto\tau^z$) and states with IVC
order ($h^{(\nu)\prime}_\mathrm{MF}\propto\tau^{x,y}s^\mu$, $\mu={0,x,y,z}$),
the latter set of which may be treated on equal footing at the level of the
$\mathrm{U}(2)_K\times\mathrm{U}(2)_{K'}$ symmetric theory. Notably, the VP state is 
clearly
hostile to the development of superconductivity. Above, we described how time reversal imposes the requirement that the (SOC-free, symmetry-unbroken) band structure energies  satisfy $\epsilon_K(\vk)=\epsilon_{K'}(-\vk)$; such resonance conditions constitute a strong prerequisite to the formation of superconductivity. Breaking time reversal symmetry and polarizing the bands according to valley thus precludes the possibility of superconductivity except in certain exotic theoretical scenarios.
Conversely, the IVC ordered states present no obvious impediment to superconductivity.

In reality, the SOC-free theory is not invariant under the full $\mathrm{U}(2)_K\times\mathrm{U}(2)_{K'}$, but instead under $\mathrm{U}(1)_c\times \mathrm{U}(1)_v\times\mathrm{SU}(2)$.
Working from the perspective of the physical $\mathrm{U}(1)_c\times \mathrm{U}(1)_v\times\mathrm{SU}(2)$ symmetry, the IVC ground state can be grouped into two categories: IVC singlets that break only the $\mathrm{U}(1)_v$ symmetry and IVC triplet states that additionally spontaneously break the spin symmetry. 
The former state is represented by mean field Hamiltonians composed of matrices $\tau^x$ and $\tau^y$, whereas the latter triplets case follows from presence of matrices $\tau^xs^{x,y,z}$ and $\tau^ys^{x,y,z}$.
Again, the large variation in Pauli violation ratio as a function of filling and our working assumption that the interacting ground state remains largely unmodified across the $\mathrm{FP}(2,2)_+$ phase (temporarily) disqualifies the triplet orders as viable candidates, leaving the IVC singlet polarized state as the proposed superconducting parent state.

On these phenomenological grounds, provided scenario (1) holds, we conclude that the addition of SOC increases the energy of the VP $\mathrm{FP}(2,2)_+$ state relative to an IVC singlet ordered state, establishing the latter as the new, SOC-mediated ground state.
A simple schematic of the energy levels as a function of $\lambda_I$ is shown in \prettyref{fig:Fig4}f.

Theoretically, the above line of reasoning can be supported on an intuitive level.
We start by making some natural assumptions regarding the nature of the interacting theory in this doping regime.
We suppose first that interactions dominate the energy scales of the problem and that these interactions necessarily favour the formation of a singly polarized state.
As addressed at the end of SI, section~\ref{theory: Polarized_phases}, the IVC singlet order parameters, e.g., $\tau^x$, anticommutes with the Ising term, $\tau^zs^z$.  
Hence, a mean field Hamiltonina $h^{(\nu)\prime}_\mathrm{MF}=\Delta_\mathrm{IVC} \tau^x/2$ still results in a singly polarized state even when $h_\mathrm{SOC}=\lambda_I\tau^zs^z$ is included.
Together, they induce an energetic separation $\sqrt{\Delta_\mathrm{IVC}^2+\lambda_I^2}$ between the flavours. 
Treating $\lambda_I$ as a perturbation to the IVC ground state and expanding in $\lambda_I/\Delta_\mathrm{IVC}$, it's clear that the change in ground state energy induced by Ising SOC will be suppressed by a factor of IVC order $\Delta E_\mathrm{MF}^\mathrm{IVC}\sim \lambda_I^2/\Delta_\mathrm{IVC}$.
By contrast, the VP mean field Hamiltonian $h_\mathrm{MF}^{(\nu)\prime} = \Delta_\mathrm{VP}\tau^z/2$ commutes with the Ising SOC Hamiltonan $h_\mathrm{SOC}=\lambda_I\tau^zs^z$. 
The introduction of SOC then splits the energy of the hitherto twofold spin degenerate energies in either valley: $\epsilon_{K^{(\prime)}}\to \epsilon_{K^{(\prime)}}\pm \lambda_I/2$.
Not only is $\Delta_\mathrm{VP}$ unable to suppress the change in mean field ground state energy, but, unlike for the IVC case, the Ising SOC has destroyed the singly polarized nature of the ground state.
We infer then that the change in mean field energy brought by Ising SOC will be larger and more positive for the VP state than it will be for the IVC state\footnote{%
Crucial to this line of reasoning is the assumption that interactions dominate the problem; the argument given above could otherwise be turned around to argue in favour of Ising SOC abetting the development of a VP state. In particular, the ease by which the VP ground state accommodates the addition of Ising SOC makes it more agreeable to the non-interacting Hamiltonian $h_\mathrm{SOC}$ compared to the IVC ground state.}.

As a proof of concept, we numerically evaluate the effect of an Ising SOC perturbation of the VP and IVC ground state energies.
Naturally, we focus on a density regime where singly polarized states are preferred.
The first step is then to  self-consistently solve for mean field Hamiltonians with VP and IVC order for a system without SOC in the manner described in SI, section~\ref{theory: Mean_field_approximation}, obtaining $H_\mathrm{MF}^{(\nu),\mathrm{VP}}$ and $H_\mathrm{MF}^{(\nu),\mathrm{IVC}}$.
The mean field ground states $|\phi_\nu^\mathrm{VP/IVC}\rangle$ are in turn employed to calculate the SOC-free mean field energies, $E_\mathrm{MF}^{\mathrm{VP/IVC},\nu}(\lambda_I=0)=\langle\phi_\nu^{\mathrm{VP/IVC}}|H_0+H_C|\phi_\nu^\mathrm{VP/IVC}\rangle$.
We subsequently introduce Ising SOC by adding $H_\mathrm{SOC}[\lambda_I]$ to the mean field Hamiltonian, $H_0+H_\mathrm{MF}^{(\nu),\mathrm{VP/IVC}}\to H_0+H_\mathrm{MF}^{(\nu),\mathrm{VP/IVC}}+H_\mathrm{SOC}[\lambda_I]$ and solving for the mean field ground state $|\phi_\nu^\mathrm{VP/IVC}(\lambda_I)\rangle$.
Importantly, the terms $H_\mathrm{MF}^{(\nu),\mathrm{VP/IVC}}$ are identical to those obtained for the $\lambda_I=0$ calculation; this calculation is no longer self-consistent. 
Finally, the Ising-perturbed energy is then calculated in the same fashion:  $E_\mathrm{MF}^{\mathrm{VP/IVC},\nu}(\lambda_I)=\langle\phi_\nu^{\mathrm{VP/IVC}}(\lambda_I)|H_0+H_C+H_\mathrm{SOC}[\lambda_I]|\phi_\nu^\mathrm{VP/IVC}(\lambda_I)\rangle$. 
We find, as expected, that the addition of SOC \emph{increases} the mean energy for both cases $\Delta E_\mathrm{MF}^{\mathrm{VP/IVC},\nu}(\lambda_I)=E_\mathrm{MF}^{\mathrm{VP/IVC},\nu}(\lambda_I)-E_\mathrm{MF}^{\mathrm{VP/IVC},\nu}(0)>0$.
In \prettyref{exfig:theory_figure_cascade}b the difference between the change in mean field energies between the VP and IVC polarized ground states is plotted as a function $\lambda_I$ for several fillings.
In accordance with our intuition, we find $\Delta E_\mathrm{MF}^{\mathrm{IVC},\nu}(\lambda_I)-\Delta E_\mathrm{MF}^{\mathrm{VP},\nu}(\lambda_I)<0$, meaning that the energy of the VP ground state increases more with the addition of $\lambda_I$ than the IVC ground state does\footnote{%
As mentioned above, however, our model does find that the IVC ground state is consistently lower energy than the VP ground state.
}. 

Although much of the logic used above will follow through, scenario (2) is somewhat more subtle. 
First, although no longer applicable across the full $\mathrm{FP}(2,2)_+$ phase, the conclusions we reached for scenario (1) \emph{do} still hold when restricting to the high doping regime.
Namely, at the large doping end of the superconducting dome, the experiments still point to a scenario in which the addition of Ising SOC pushes the system away from a VP normal state towards an IVC singlet normal state.
Further, since an evolving SOC-mediated $\mathrm{FP}(2,2)_+$ ground state need not imply an evolving SOC-free ground state, we continue in our assumption that a VP state is realized prior to the addition of SOC.
The primary distinction between scenarios (1) and (2) is therefore that we can reject neither the SVP nor the IVC triplet phases as candidate ground states at low dopings on the basis of the small PVR at large dopings.
Among the potential IVC triplet states, the reasoning provided above for why Ising would prefer an IVC singlet over the VP state similarly selects the $z$-component IVC triplet, represented by $\tau^{x,y}s^z$, as the most likely candidate.
We are therefore left with a possible mixed mean field Hamiltonian of the form
\begin{align}\label{eqn:mixed_order_hMF}
    h_\mathrm{MF}^{(\nu)\prime}=\frac{\Delta^\mathrm{singlet}_\mathrm{IVC}(\nu)}{2}\tau^x
    + \frac{\Delta^\mathrm{triplet}_\mathrm{IVC}(\nu)}{2}\tau^ys^z
    + \frac{\delta\lambda_I(\nu)}{2}\tau^zs^z,
\end{align}
where we have explicitly indicated the functional dependence of the mean field order parameters on the filling $\nu$.
Note that the IVC orders $\tau^x$ and $\tau^ys^z$ were chosen such that they anticommute, guaranteeing that \eqref{eqn:mixed_order_hMF} describes a singly polarized phase.

We can make some arguments towards the functional form of the mean field parameters $\Delta_\mathrm{IVC}^\mathrm{singlet}(\nu)$, $\Delta_\mathrm{IVC}^\mathrm{triplet}(\nu)$, and $\delta\lambda_I(\nu)$.
First, even at the very edge of the underdoped supercondcutor, where the PVR is at its largest, a mean field Hamiltonian in which the IVC single component completely vanishes, $\Delta_\mathrm{IVC}^\mathrm{singlet}(\nu_{\mathrm{low}\;\mathrm{dopings}})=0$, remains unlikely. 
For instance, in the simulations of \prettyref{exfig:theory_figure_cascade}c,d, IVC order is prohibited and the singly polarized phases (shaded red) are characterized entirely by an interaction-induced increase $\delta\lambda_I\sim 3.5$~meV of the effective Ising SOC.
Such an extreme enhancement of the effective Ising would in turn imply a PVR of order $\sim20$.
It's therefore likely that $|\Delta_\mathrm{IVC}^\mathrm{singlet}(\nu)|\geq\sqrt{\left[\Delta_\mathrm{IVC}^\mathrm{triplet}(\nu)\right]^2+\left[\lambda_I+\delta\lambda_I(\nu)\right]^2}$ throughout the $\mathrm{FP}(2,2)_+$ phase.

We can similarly discuss the relative magnitudes of $\delta\lambda_I$ and $\Delta_\mathrm{IVC}^{\mathrm{triplet}}$.
On the one hand, the SOC term $h_\mathrm{SOC}$ very clearly prefers the development of an Ising SVP phase, suggesting that $\delta\lambda_I$ may be the next-largest contribution to $h^{(\nu)\prime}_\mathrm{MF}$ after $\Delta_\mathrm{IVC}^{\mathrm{singlet}}$.
Such an interplay between the interaction-induced IVC singlet and  Ising orders is in fact seen in the numerics presented in \prettyref{exfig:theory_figure_cascade}c,d: $\Delta_\mathrm{IVC}^{\mathrm{singlet}}(\nu)$ increases with filling whereas $\delta\lambda_I(\nu)$ decreases\footnote{%
We acknowledge that our simulations also find an overly-large mean-field enhancement of the Ising SOC even in the presence of IVC.
For the $C_3$ symmetry simulation shown in \prettyref{exfig:theory_figure_cascade}e, the parameters obtained in the low-doping region of the $\mathrm{FP}(2)_+$ phase, $\nu_{\mathrm{low}\;\mathrm{doping}}\sim -6\times10^{11}\text{ cm}^{-2}$, are given by $\delta\lambda_I(\nu_{\mathrm{low}\;\mathrm{doping}})\sim3$~meV, $\Delta_\mathrm{IVC}^{\mathrm{singlet}}(\nu_{\mathrm{low}\;\mathrm{doping}})\sim2$~meV. 
Towards the large doping end of the IVC ordered region, $\nu_{\mathrm{large}\;\mathrm{doping}}\sim-7\times10^{11}\text{ cm}^{-2}$, the relative magnitudes are the IVC and Ising mean field parameters are interchanged: $\delta\lambda_I(\nu_{\mathrm{large}\;\mathrm{doping}})\sim2.5$~meV, $\Delta_\mathrm{IVC}^{\mathrm{singlet}}(\nu_{\mathrm{large}\;\mathrm{doping}})\sim2.7$~meV.
Hence, although the numerics demonstrate the correct trends, they again fail to quantitatively account for the experimental observations.}.
Conversely, the symmetry arguments above make it very natural for the IVC singlet and IVC $z$-triplet orders to have very similar energies. 
In fact, the degeneracy between the two orders, $\tau^{x,y}$ and $\tau^{x,y}s^z$, is only lifted at the level of the valley Hund's interaction and Rashba energy scale.
Indeed the introduction of Ising SOC breaks the large symmetry group $\mathrm{U}(2)_K\times\mathrm{U}(2)_{K'}$ down to $\mathrm{U}(1)_c\times\mathrm{U}(1)_v\times\mathrm{U}(1)_{K,z}\times\mathrm{U}(1)_{K',z}$, under whose action the singlet and $z$-triplet IVC orders are still able mix.

\section{Orbital coupling to in-plane magnetic fields}
\label{theory: Orbital_coupling}

Magnetic fields oriented in the graphene plane also enter the low-energy theory of BLG through orbital effects, i.e., a renormalization of the hopping terms due to the magnetic flux between the two graphene layers. While this effect vanishes for purely two-dimensional monolayer graphene, in BLG the orbital coupling scales linearly with the finite width $d$ between the layers. The leading-order contribution of this type comes from the renormalization of the intralayer nearest-neighbor hopping term $\gamma_0$, which is the largest energy scale in the problem by an order of magnitude. Choosing a gauge that preserves translation invariance in the plane, $\vA = d (B_y, -B_x, 0)$ where $d = 0.33$ nm is the interlayer distance, this leads to orbital contributions on layers 1 and 2 given by\cite{kheirabadiMagneticRatchetEffect2016}
\begin{align}
    h_{\rm orbital, 1} = v_0 \left( - \tau_z \sigma_x b_y + \sigma_y b_x \right) \\
    h_{\rm orbital, 2} = v_0 \left( \tau_z \sigma_x b_y - \sigma_y b_x \right)    
\end{align}
where $b_j = e d B_j/2 \hbar$ and $j=x,y$. The system thus has an orbital magnetic moment given (to leading order) by 
\begin{equation}
    \mu_{\rm orbital} = \frac{v_0 e d}{2 \hbar} = \frac{\sqrt{3} \gamma_0 a_0 e d }{4 \hbar} \approx 0.14 \frac{\rm meV}{\rm Tesla} .
\end{equation}
This scale nominally yields a stronger coupling than the spin Zeeman term $h_{\rm Z} = \mu_B \vB \cdot \vs$ with the Bohr magneton $\mu_B \sim 0.06$ meV/Tesla. However, orbital effects will be suppressed near the $\vK, \vK'$ points because of the strong sublattice polarization of the low-energy wavefunctions---analogous to the suppression of the ``bare" Rashba SOC discussed above.

Projecting down to
the $2\times2$ low-energy subspace spanned by the $A1$ and $B2$ sites enables estimation of the effective orbital coupling, given to leading order as\cite{kheirabadiMagneticRatchetEffect2016}
\begin{equation}
    h^\xi_{\rm orbital}(\vk) = \frac{2 v_0^2}{\gamma_1^2} u \left( \vk \times \vb \right)_z \sigma_0 s_0 = \mu^{\rm eff}_{\rm orbital} B_x \sigma_0 s_0,
    \label{eq:orbital_coupling_projected}
\end{equation}
where $\vk$ is the momentum measured from either of the Dirac points $\vK, \vK'$. We picked the in-plane field in the $x$ direction and defined the effective orbital magnetic moment as
\begin{equation}
\mu^{\rm eff}_{\rm orbital} = - \frac{3 a_0 e d \gamma_0^2}{4 \hbar \gamma_1^2} u (k_y a_0).
\end{equation}
For $u \approx -80$ meV (corresponding to positive $D \approx 1$ V/nm) one finds $\mu^{\rm eff}_{\rm orbital} \sim 0.4 \frac{ \rm meV}{\rm Tesla} (k_y a_0)$, which for Fermi momenta satisfying $k_y a_0 = k_F a_0 \sim 0.05$ near the center of the small pockets gives $\mu^{\rm eff}_{\rm orbital} \sim 0.02 \frac{ \rm meV}{\rm Tesla}$, or $\mu^{\rm eff}_{\rm orbital} = 0.35 \mu_B$. In the following discussion we will use the dimensionless quantity $g_0 k_F$ to denote the strength of the orbital coupling, defined as $\mu^{\rm eff}_{\rm orbital} = g_0 k_F \mu_B$.

\section{BCS mean-field analysis in the presence of Zeeman field, SOC terms and orbital coupling}
\label{theory: BCS_meanfield}

In this section, we describe the model used to investigate the evolution of the Pauli-limit violation ratio in our sample, as well as its analytical solution. We follow the treatment first developed in Ref.~\citenum{frigeriSuperconductivityInversionSymmetry2004} to compute the response to a Zeeman field of non-centrosymmetric superconductors with Rashba SOC, later generalized in Refs.~\citenum{luEvidenceTwodimensionalIsing2015, saitoSuperconductivityProtectedSpinvalley2016} to systems with a coexistence of Ising and Rashba SOC. We also incorporate the effect of orbital depairing in a simple model for BLG inspired by  Ref.~\citenum{kheirabadiMagneticRatchetEffect2016} and the discussion in SI, section~\ref{theory: Orbital_coupling}.

We assume that as a result of a symmetry-breaking transition (cascade), the system is in an $\mathrm{FP}(2,2)_{+}$ phase with two large and two small hole pockets.   
We model the small pockets by two electronic bands centered around trigonal-warping loci $\pm \vec{T}$ that respectively originate from the $K$ and $K'$ valleys. In other words, we imagine a scenario where exchange interaction effects promote nematic order such that electrons in each valley ``flock" from evenly occupying the three small pockets to completely polarizing one pocket\cite{dongIsospinFerromagnetismMomentum2021}---and further that the selected pockets in the two valleys are time-reversed partners of each other. We note that the quantum oscillation data do not directly reveal nematicity, but do indicate the presence of only two small Fermi pockets (instead of the 6 pockets predicted by the non-interacting band structure). The following modelling could also apply to a situation where the %
$\mathrm{FP}(2,2)_{+}$ state is not nematic, provided that time-reversal symmetry $\mathcal{T}$, which relates the two remaining small pockets, is preserved.

Adopting the preceding scenario, we take the normal-state Hamiltonian to be
\begin{equation}
    H(\xi \vec{T} + \vk) = \xi_{\vk}  + \frac{1}{2} \xi g_I s^z + \frac{1}{2} g_R(\vec{s}\times\vec{k})\cdot \vec{z} + \vb  \cdot  \vs + g_{\rm orb} \left( \vb \times (\vec{k} + \vec{k}_0) \right) \cdot \vec{z}    ,
    \label{eq:toy_model}
\end{equation}
with $\xi = \pm 1$ the valley index, $\vec{T}$ the momentum of one of the trigonal-warping loci, and ${\vec s} = (s^x, s^y, s^z)$ a vector of Pauli matrices that act on the spin degree of freedom. On the right side $\xi_{\vk}$ is the spin-orbit-free normal state band structure, which we linearize near the Fermi surface as $\xi_{\vk} \approx v_F (k-k_F)$ ($k_F$ denotes the Fermi momentum measured from the center $\vec{T}$ of the pocket). The next two terms incorporate Ising and Rashba SOC with strengths $g_I$ and $g_R$, respectively.  
The final two terms incorporate effects of an in-plane magnetic field $B_\parallel$, packaged into a vector $\vb = (\mu_B B_\parallel, 0, 0)$ with $\mu_B$ the Bohr magneton: $\vb \cdot \vs$ is simply the Zeeman energy while $g_{\rm orb}$ captures orbital effects\cite{kheirabadiMagneticRatchetEffect2016} of the in-plane field. The momentum shift $\vk_0 = \vec{T}-\vec{K}$ captures the fact that the relevant momenta for orbital effects of in-plane fields are measured with respect to the Dirac points rather than the center $\vec{T}$ of the small pockets.

We then consider a local (momentum-independent) spin-singlet pairing term that gives rise to superconductivity with a critical temperature $T_c^0$ at zero magnetic field. In the presence of an in-plane magnetic field $B_\parallel$ the superconductivity is weakened through a combination of spin and orbital effects, with $T_c < T_c^0$ given by the solution of a self-consistent gap equation, linearized near the second-order transition at $T_c$ where the pairing amplitude $\Delta \rightarrow 0$,\cite{saint-jamesTypeIISuperconductivity1969, frigeriSuperconductivityInversionSymmetry2004, luEvidenceTwodimensionalIsing2015,saitoSuperconductivityProtectedSpinvalley2016, zwicknaglCriticalMagneticField2017}
\begin{equation}
    \ln \left( \frac{T_c}{T_c^0} \right) =  \frac{T_c}{2} \sum_{\omega_n}\left(\left\langle \int d \xi_{\vk} \mathrm{Tr}\left\{s_y G_0(\vec{T}+\vec{k},i\omega_n) s_y G_0^*(-\vec{T}-\vec{k}, i\omega_n)\right\}\right>_{\rm FS}-\frac{\pi}{|\omega_n|}\right) .
\end{equation}
Here $\sum_{\omega_n}$ denotes a summation over Matsubara frequencies $i\omega_n$, $\langle\cdots\rangle_{\rm FS}$ denotes a Fermi surface average and $\mathrm{Tr}\{ \cdots \}$ is a trace over spin Pauli matrices, and $G_0(\vec{T}+\vec{k}, i\omega_n)$ is the normal-state Green's function given by
\begin{equation}
    G_0(\vec{T}+\vec{k},i \omega_n) = \frac{(i\omega_n-\chi_+)+\vec{p}_+ \cdot \vs}{(i \omega_n-\chi_+)^2-\vec{p}_{+}^2}\,,\quad  G_0^*(-\vec{T}-\vec{k},i \omega_n) = \frac{(-i\omega_n-\chi_-)+\vec{p}_- \cdot \vs^*}{(-i \omega_n-\chi_-)^2-\vec{p}_{-}^2} .
\end{equation}
For convenience we introduced $\chi_\pm = \xi_{\vec{k}}\pm g_{\rm orb} (k_y+k_{0,y}) \mu_B B_\parallel$ ($k_{0,y}$ is the $y$-component of $\vec{k}_0$) and $\vec{p}_{\pm}= \left(  \pm g_R k_y/2 + \mu_B B_\parallel, \mp g_R k_x/2, \pm g_I/2 \right)$. Carrying out the $\xi_{\vec{k}}$ integral, Fermi surface average, and Matsubara summation yields a final form of the gap equation:
\begin{equation}
    \ln \left( \frac{T_c}{T_c^0} \right) + \Phi(\rho_-, \tilde{\chi}_0) + \Phi(\rho_+,\tilde{\chi}_0) - \frac{\tilde{\vec{p}}_+\cdot \tilde{\vec{p}}_-}{|\tilde{\vec{p}}_+||\tilde{\vec{p}}_{-}|} \left[\Phi(\rho_-,\tilde{\chi}_0) - \Phi(\rho_+,\tilde{\chi}_0) \right]= 0,
    \label{eq:self_consistent_result}
\end{equation}
where $\tilde{\chi}_0 = -\tilde{g}_{\rm orb} \mu_B B_\parallel / 2 \pi T_c$ ($\tilde{g}_{\rm orb}\equiv g_{\rm orb} (k_F+k_{0,y})$ denotes a characteristic scale for the orbital depairing), $\tilde{\vec{p}}_{\pm} = (\pm g_R k_F/2 + \mu_B B_\parallel, \mp g_R k_F/2, \pm g_I/2)$, and $\rho_\pm = (|\tilde{\vec{p}}_+| \pm |\tilde{\vec{p}}_-|)/2\pi T_c$. The function $\Phi(\rho,\tilde{\chi}_0)$ is defined in terms of the digamma function $\psi(z)$ as
\begin{equation}
    \Phi(\rho,\tilde{\chi}_0)=\frac{1}{4}\left\{\mathrm{Re}\left[\psi\left(\frac{1+i\rho}{2}+i \tilde{\chi}_0\right)-\psi\left(\frac{1}{2}\right)\right]+\mathrm{Re}\left[\psi\left(\frac{1+i\rho}{2}-i \tilde{\chi}_0 \right)-\psi\left(\frac{1}{2}\right)\right]\right\}\,.
\end{equation}
In the limit $\tilde{\chi}_0 \to 0$, i.e., without orbital depairing, Eq.~\eqref{eq:self_consistent_result} reduces to the form used in Ref.~\citenum{luEvidenceTwodimensionalIsing2015}. This equation can be solved numerically to obtain the relationship between the critical temperature $T_c$ and the critical in-plane field $B_{c \parallel}$ of the superconductor,
given input SOC and orbital coupling parameters. As in the main text we denote the limits of zero-field critical temperature and zero-temperature critical field by $T_c^0$ and $B_{c \parallel}^0$, respectively.

\prettyref{exfig:theory_figure_PLV}a,b displays the dependence of the Pauli-limit violation ratio on the Ising SOC ($g_I$), Rashba SOC ($g_R$) and orbital coupling ($\tilde{g}_{\rm orb}$) parameters. In the presence of a purely Ising-type SOC ($g_R = \tilde{g}_{\rm orb} = 0$), the evolution of $B_{c \parallel}$ as a function of $T_c$ and $g_I$ shows the characteristic low-temperature divergence\cite{frigeriSuperconductivityInversionSymmetry2004} of
$B_{c \parallel}$, which is due to the inability of the in-plane field---perpendicular to the Ising spin quantization axis---to destroy the resonance condition for spin-singlet pairing between electronic states at $\vk$ and $-\vk$. In other words, there is always a non-zero density of electronic states available for pairing opposite spin-components (albeit decreasing with $B_\parallel$), thus leading to persistent superconductivity\footnote{Our linearized treatment can only capture second-order transitions, and thus neglects a possible first-order transition to a polarized normal state, in analogy with van Vleck paramagnetism. In this framework, the Zeeman energy gain in the normal state scales as $E_z \sim \left( \mu_B B \right)^2/\lambda_I$ to second order in perturbation theory, instead of the standard form $\mu_B B$. This is due to the spin projection being locked in the out-of-plane directions by Ising SOC. Equating to the condensation energy $\Delta$ leads to the condition $B_c \sim \sqrt{\Delta \lambda_I /\mu_B}$ or $B_c / B_p \sim \sqrt{\lambda_I/\Delta}$ with the Pauli limiting field $B_p = \Delta_0/\sqrt{2} \mu_B$, a scaling form often quoted in the study of Ising superconductivity in TMDs~\cite{luEvidenceTwodimensionalIsing2015, saitoSuperconductivityProtectedSpinvalley2016, xiIsingPairingSuperconducting2016}} as $T \rightarrow 0$. In contrast, when either Rashba SOC\footnote{In the case of Rashba SOC we recover the $\sqrt{2}$ enhancement of the critical field compared to a spin-degenerate metallic state, first predicted by Gork'ov and Rashba\cite{gorkovSuperconducting2DSystem2001} based on a microscopic calculation of the in-plane spin susceptibility.} or orbital decoupling is added, the low-temperature divergence of $B_{c \parallel}$ is strongly suppressed, alongside a reduction of the Pauli-limit violation at all temperatures. This suppression occurs because both Rashba and orbital effects lead to a non-trivial deformation of the Fermi pockets under an in-plane field, which destroys the resonance conditions necessary for spin-singlet, zero-momentum pairing (we neglect finite-momentum, FFLO-type pairing channels in our analysis).

\section{Modeling of the Pauli-limit violation data}
\label{theory: PVR_fit}

We now describe various efforts at fitting the Pauli-limit violation data from \prettyref{fig:Fig4}c and \prettyref{exfig:inplanefield}f to our theoretical model. The difference between these two data sets is that $B_{c \parallel}$ reported in \prettyref{fig:Fig4}c is obtained by a phenomenological extrapolation to zero temperature, whereas $B_{c \parallel}$ in \prettyref{exfig:inplanefield}f is measured at a fixed base temperature $T_c \sim 30$ mK. 

In the following we fix the value of Ising SOC to $g_I = 0.7$ meV as extracted from quantum Hall measurements, and assume that it remains constant as a function of doping (thus ignoring a potential interaction-induced enhancement of its bare value). We first consider Rashba and orbital effects separately. Model 1 (see Table below) considers only Rashba SOC ($\tilde{\mathrm{g}}_{\rm orb} = 0$), which because of its linear dependence on $k_F$ is expected to scale as $g_R k_F \sim \sqrt{n_{\rm small}}$ with $n_{\rm small}$ the electronic density in the small pocket. Due to the difficulty of extracting the size of the small Fermi pockets directly from quantum oscillation data, we take a phenomenological ansatz for the small pocket density. Motivated by the quantum oscillation data for the ${\rm FP}(2,2)_+$ phase, we assume that the normalized SdH frequency $f_\nu = n_{\rm small}/n$ ($n$ is the total electronic density in the system) varies linearly with $n$ in the ${\rm FP}(2,2)_+$ phase---that is, we take $f_\nu = a n + b$ with constants $a$, $b$ to be treated as fitting parameters. This assumption leads to the form $g_R k_F \sim \sqrt{an^2 + bn}$ for the energy scale associated with Rashba coupling.

As shown in \prettyref{exfig:theory_figure_PLV}c,d, Model 1 fits the PVR data well. However we find that to account for the strong PVR dependence, the Rashba energy scale $g_R k_F$ must vary strongly over the corresponding doping range. In the current Model 1, this variation occurs through an order-of-magnitude increase in the electronic density in the small pocket, which would require the small pocket density to nearly vanish at the edge of the SC dome. 
Part of the variation in $g_R k_F$ could also come from a doping dependence of the Rashba SOC parameter $g_R$ itself, either due to interaction-induced renormalization or through band structure effects (e.g., higher-order terms in the low-energy projection that relates the bare $\lambda_R$ to $g_R$ in the low-energy description, Eq.~\ref{eq:toy_model}).

The same conclusion is obtained for the case of only orbital effects ($g_R=0$,
see fits to the data in \prettyref{exfig:theory_figure_PLV}c,d). A model with
only an orbital source of depairing (as introduced in the previous section) can
similarly account for the suppression of PVR as hole doping is increased. In
our fits, motivated by the form of Eq.~\eqref{eq:toy_model}, we assume for
simplicity a dependence of $\tilde{g}_{\mathrm{orb}}\sim a\sqrt{n}+b$ where the
constant offset qualitatively stems from the finite location of the locus of
the trigonally warped pockets with respect to the Dirac point. In Models 2, 3
and 4 (see Table at the end of this section) we assume: filling-dependent $g_R
k_F$ and constant orbital coupling $\tilde{g}_{\mathrm{orb}} \sim c$; $g_R k_F
= 0$ and filling dependent $\tilde{g}_{\mathrm{orb}}$; and both filling
dependent $g_R k_F$ and $\tilde{g}_{\mathrm{orb}}$.
 
In summary, our theoretical modeling can account for the observed PVR evolution
through a strong doping-dependence of either Rashba SOC and/or orbital
depairing effects, both of which compete with Ising SOC. This is reminiscent to
the phenomenology of Ref.~\citenum{luEvidenceTwodimensionalIsing2015}, where a
doping-dependent PVR was observed and attributed to increasing effects of
Rashba SOC with carrier density. However, the required doping dependence
appears very large in view of naive band-structure estimates. Consequently,
band structure reconstruction mediated by electron interactions in the ${\rm
FP}(2,2)_+$ phase must be significant for our scenario to capture the physics
responsible for the evolution of Pauli-limit violation---providing a guidance
for further theory modelling of superconductivity in BLG that is beyond the
scope of this work. A potential resolution of this issue could involve an
interaction-induced enhancement of Ising SOC---or the nucleation of another
symmetry-broken phase which increases the out-of-plane spin canting, see
discussion below Eq.~\eqref{eqn:mixed_order_hMF}---in a doping dependent
manner, particularly near the low hole doping region of the ${\rm FP}(2,2)_+$
phase.
 
\begin{table}
\begin{center}
    
\def\arraystretch{1.5}
\begin{tabular}{|c|c|c|c|c|l|l|}
\hline
Model \# & $g_I$ (meV) & $g_R k_F$ (meV) & $\tilde{g}_{\rm orb}$ (1) & $T=30$ mK & $T\to 0$ \\
\hline
1       & $0.7$ & $\sqrt{ a n + b n^2  }$ & $0$ &\begin{tabular}{@{}l@{}}$a = -0.212 $ \\ $b = 0.0359$\end{tabular} &  \begin{tabular}{@{}l@{}}$a = -0.137$ \\ $b = 0.0231$\end{tabular}\\
\hline
2       & $0.7$ & $\sqrt{ a n + b n^2  }$ & $c$ & \begin{tabular}{@{}l@{}}$a = -0.225 $ \\ $b = 0.0382$ \\ $c = 0.0848$\end{tabular} &  \begin{tabular}{@{}l@{}}$a = -0.136 $ \\ $b = 0.0236$ \\ $c = 0.205$\end{tabular}\\
\hline
3       & $0.7$ & 0 & $a \sqrt{n} + b$ & \begin{tabular}{@{}l@{}}$a = 1.81 $ \\ $b = -4.25$\end{tabular} &   \begin{tabular}{@{}l@{}}$a = 1.79 $ \\ $b = -4.24$\end{tabular}\\
\hline
4       & $0.7$ &  $\sqrt{ a n + b n^2  }$ & $c \sqrt{n} + d$ & \begin{tabular}{@{}l@{}}$a = -0.259 $ \\ $b = 0.0439$ \\ $c = 0.0122$ \\ $d = 0.0872$\end{tabular} &  \begin{tabular}{@{}l@{}}$a = -0.134 $ \\ $b = 0.0227$ \\ $c = 0.0426$ \\ $d = -0.0796$\end{tabular}\\
\hline

\end{tabular}

\end{center}

\caption{Models used for the fitting procedure. Here $n$ is the total doping
density in units of $\times 10^{11} ~\text{cm}^{-2}$. Units of $a$, $b$, $c$,
$d$ fitting parameters are chosen to yield correct units of the physical
parameters $g_I$, $g_R k_F$, $\tilde{g}_{\rm orb}$. Their filling dependence is
shown in \prettyref{exfig:theory_figure_PLV}e,f. } 

\end{table}

\end{document}